 \documentclass[preprint,review,12pt,sort&compress]{elsarticle}
\usepackage[margin=1.18 in]{geometry}
\usepackage{multirow}
\newcommand{\B}{Ba\v zant}  

 \newcommand{\bc}{\begin{center}}
 \newcommand{\ec}{\end{center}}
                   \newcommand{\bfr}{\begin{flushright}}
                   \newcommand{\efr}{\end{flushright}}
   \newcommand{\ii}{\item}
     \newcommand{\be}{\begin{enumerate}}
     \newcommand{\ee}{\end{enumerate}}
        \newcommand{\bi}{\begin{itemize}}
        \newcommand{\ei}{\end{itemize}}
            \newcommand{\bd}{\begin{description}}
            \newcommand{\ed}{\end{description}}
                \newcommand{\beq}{\begin{equation}}
                \newcommand{\eeq}{\end{equation}}
                  \newcommand{\bea}{\begin{eqnarray}}
                  \newcommand{\eea}{\end{eqnarray}}

      \newcommand{\bfi}{\begin{figure}}
      \newcommand{\efi}{\end{figure}}
\newcommand{\bay}{\begin{array}{l}}
\newcommand{\eay}{\end{array}}
            
\usepackage{amsmath}
\usepackage{breqn}
\usepackage{textcomp}

\usepackage{amssymb}

\journal{Composite Structures}

\begin{document}
\begin{titlepage}

\clearpage\thispagestyle{empty}



\noindent

\hrulefill

\begin{figure}[h!]

\centering

\includegraphics[width=2 in]{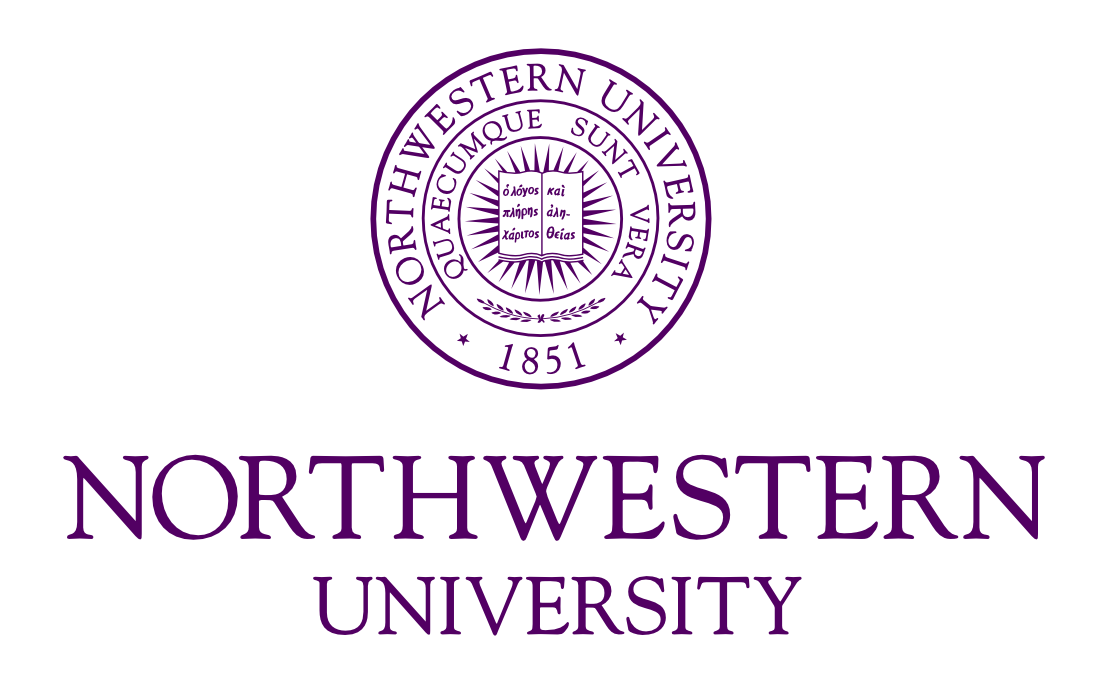}

\end{figure}


\begin{center}


{\bf Center for Sustainable Engineering of Geological and Infrastructure Materials (SEGIM)} \\ [0.1in]

Department of Civil and Environmental Engineering \\ [0.1in]

McCormick School of Engineering and Applied Science \\ [0.1in]

Evanston, Illinois 60208, USA



\end{center} 

\hrulefill \\ \vskip 2mm

\vskip 0.5in

\begin{center}

{\large {\bf SPECTRAL STIFFNESS MICROPLANE MODEL FOR QUASIBRITTLE TEXTILE COMPOSITES}}\\[0.5in]

{\large {\sc Marco Salviato, Shiva Esna Ashari, Gianluca Cusatis}}\\[0.75in]

{\sf \bf SEGIM INTERNAL REPORT No. 15-09/707S}\\[0.75in]

\end{center}

\noindent {\footnotesize {{\em Submitted to Composite Structures \hfill September 2015} }}

\end{titlepage}

\newpage

\clearpage \pagestyle{plain} \setcounter{page}{1}
\begin{frontmatter}



\begin{abstract}
\linespread{1}\selectfont
The present contribution proposes a general constitutive model to simulate the orthotropic stiffness, pre-peak nonlinearity, failure envelopes, and the post-peak softening and fracture of textile composites.

Following the microplane model framework, the constitutive laws are formulated in terms of stress and strain vectors acting on planes of several orientations within the material meso-structure. The model exploits the spectral decomposition of the orthotropic stiffness tensor to define orthogonal strain modes at the microplane level. These are associated to the various constituents at the mesoscale and to the material response to different types of deformation. Strain-dependent constitutive equations are used to relate the microplane eigenstresses and eigenstrains while a variational principle is applied to relate the microplane stresses at the mesoscale to the continuum tensor at the macroscale.

Thanks to these features, the resulting spectral stiffness microplane formulation can easily capture various physical inelastic phenomena typical of fiber and textile composites such as: matrix microcracking, micro-delamination, crack bridging, pullout, and debonding. The application of the model to a twill 2$\times$2 shows that it can realistically predict its uniaxial as well as multi-axial behavior. Furthermore, the model shows excellent agreement with experiments on the axial crushing of composite tubes, this capability making it a valuable design tool for crashworthiness applications.

The formulation is computationally efficient, easy to calibrate and adaptable to other kinds of composite architectures of great current interest such as 2D and 3D braids or 3D woven textiles.

\end{abstract}




\end{frontmatter}


\section{Introduction}
\label{}
Thanks to their excellent specific mechanical performances and the recent developments in manufacturing technologies, the range of engineering applications of textile composites is continuously expanding. Current applications include land, marine and air transportation, wind and tidal energy production, and blast protection of civil infrastructures and vehicles \cite{Cho92,DanIsh92,BogPas96}. However, in order to take advantage of the outstanding characteristics of these materials, design tools to simulate the orthotropic stiffness, pre-peak nonlinearity, failure envelopes, and the post-peak softening and fracture are quintessential.

Since the pioneering works by Ishikawa and Chou \cite{IshCho82,IshCho83} and Ishikawa \emph{et al.} \cite{IshMatHay85}, several formulations have been proposed, with varying degrees of success, to model the elastic properties of textile composites \cite{ZhaHar90,NaiGan92, HahPan94, CoxCarFle94,SciAboBen97,LomHuyLuo01,LomIvaVer07} and their failure mechanisms \cite{BlaWalHan93,Nai95,SciAboBen99,Hua00,CarPog01,NicRiv04,ProLomVer11}. In general, however, these models stand on strength criteria to describe failure of the mesoscale constituents thus lacking completely of any description of the fracture mechanics involved. This is a serious deficiency being extensive intra-laminar cracking one of the main failure mechanisms in most applications of textile composites.

Modeling the fracturing behavior of textile composites, not only requires a fracture mechanics framework, it also urges the acknowledgment of their \emph{quasi-brittle} character which highly affects the process of crack nucleation and growth. In facts, due to the complex mesostructure characterizing quasi-brittle materials (such as composites and nanocomposites, ceramics, rocks, sea ice, bio-materials and concrete, just to mention a few), the extent of the non-linear Fracture Process Zone (FPZ) occurring in the presence of a macrocrack is usually not negligible \cite{BazPla98}. The stress field along the FPZ is nonuniform and decreases with crack opening gradually, due to discontinuous cracking, crack bridging by fibers, and frictional pullout of inhomogeneities. As a consequence, the fracturing behavior and, most importantly, the energetic size effect and the quasibrittleness effect associated with structure geometry, cannot be described by means of the classical Linear Elastic Fracture Mechanics (LEFM).
To capture the effects of a finite FPZ size, the introduction in the formulation of a characteristic (finite) length scale of the material is necessary \cite{BazDanLi96,BazPla98}. This is attempted in the present work.

Inspired by a recent theoretical framework for unidirectional composites by Cusatis \emph{et al.} \cite{CusBegBaz08,BegCusBaz08}, this contribution aims at proposing a general constitutive model to simulate the damaging and fracturing behavior of textile composites. The formulation stands on the definition of strain-dependent constitutive laws in terms of stress and strain vectors acting on planes of several orientations within the material meso-structure. In this way, the model can easily capture various physical inelastic phenomena typical of fiber and textile composites such as: matrix microcracking, micro-delamination, crack bridging, pullout, and debonding.

Thanks to the coupling with the crack band model \cite{Baz76,BazOh83}, the formulation is endowed with a characteristic length dependent on the strength and the fracture energy of the material. This is key to capture the intra-laminar size effect, a salient feature of composite structures. This aspect, too often overlooked in the literature on composites, is a determinant factor for damage tolerance design of large composite structures.

\section{Theoretical framework}
\subsection{Microplane model}
Inspired by the slip theory of plasticity pioneered by Taylor \cite{Tay38} and later refined by Batdorf and Budiansky \cite{BatBud49}, the microplane theory was originally developed to describe the softening damage of heterogeneous but statistically isotropic materials such as concrete and rocks \cite{Baz84, BazOh85}. Since its introduction in the early 1980s, the microplane model for concrete has evolved through 7 progressively improved versions labeled as M1 \cite{Baz84, BazOh85}, M2 \cite{BazPra88}, M3 \cite{BazXiaPra96}, M4 \cite{BazCanCar00,BazDil04}, M5 \cite{BazCan05}, M6 \cite{CanBaz11}, M7 \cite{CanBaz13} and it has been recently adopted for the simulation of concrete at early age \cite{DilCus13}. Microplane models have also been developed for
other complex materials such as jointed rock \cite{CheBaz14}, sand, clay, rigid foam,
shape memory alloys, and unidirectional and textile composites \cite{BazPra87,BroBaz00,BroBazDan01,CusBegBaz08,BegCusBaz08,CanBazHoo11,KirSalBaz15}. A high order microplane model \cite{CusZho13} was also derived recently on the basis of an underlying discrete model \cite{CusPelMen11,CusMenPel11}.

A key feature of the microplane model is that the constitutive laws are formulated in terms of the stress and strain vectors acting on a generic plane of any orientation within the material meso-structure, called the \textit{microplane}. These planes can
be conceived as the tangent planes of a unit sphere surrounding every point in the three-dimensional space (Fig. \ref{f1}a). The microplane strain
vectors are the projections of the macroscropic strain tensor, whereas the macroscopic stress tensor is related to the microplane stress vectors via
the principle of virtual work. The adoption of vectors rather than tensors makes the approach conceptually clearer while the introduction of microplanes allows to inherently embed the effect of the mesostructure into the formulation.

In this contribution, a kinematically constrained microplane model is adopted. This means that the strain vector on each microplane is the projection of the macroscopic strain tensor. In \emph{kelvin notation} \cite{Thomson78,Elbing94}
this reads:
 \begin{equation}\label{e1}
 \boldsymbol{\varepsilon}_P=\boldsymbol{\mathcal{P}}\boldsymbol\varepsilon
 \end{equation}
where $\boldsymbol\varepsilon_P=\left[\varepsilon_N~ \varepsilon_M~ \varepsilon_L \right]^T$ represents the microplane strain vector (Fig \ref{f1}a) with
$\varepsilon_N=$ normal strain component and $\varepsilon_M$ and $\varepsilon_L=$ shear strain components. Further,
 \begin{equation}\label{e2}
\boldsymbol{\mathcal{P}}=
\begin{bmatrix}
N_{11} & N_{22} & N_{33} & \sqrt{2}N_{23} & \sqrt{2}N_{13} & \sqrt{2}N_{12} \\
M_{11} & M_{22} & M_{33} & \sqrt{2}M_{23} & \sqrt{2}M_{13} & \sqrt{2}M_{12} \\
L_{11} & L_{22} & L_{33} & \sqrt{2}L_{23} & \sqrt{2}L_{13} & \sqrt{2}L_{12} \\
\end{bmatrix}
 \end{equation}
is a $3\times6$ matrix relating the macroscopic strain tensor to the microplane strain as a function of the plane
orientation. In facts, $N_{ij}=n_in_j$, $M_{ij}=(m_in_j+m_jn_i)/2$ and $L_{ij}=(l_in_j+l_jn_i)/2$, where
$n_i$, $m_i$ and $l_i$ are local Cartesian coordinate vectors on the generic microplane with $n_i$ being the i-$th$ component
of the normal (Fig. \ref{f1}a). With reference to the spherical coordinate system represented in Fig. (\ref{f1}b), the foregoing components can
be expressed as a function of the spherical angles $\theta$ and $\varphi$: $n_1=\sin{\theta}\cos{\varphi}$, $n_2=\sin{\theta}\sin{\varphi}$,
$n_3=\cos{\theta}$ while one can choose $m_1=\cos{\theta}\cos{\varphi}$, $m_2=\cos{\theta}\sin{\varphi}$, $m_3=-\sin{\theta}$ which gives, for orthogonality,
$l_1=-\sin{\varphi}$, $l_2=\cos{\varphi}$ and $l_3=0$.

According to the microplane framework, the constitutive laws are then defined at the microplane level in a vectorial form.
This makes the formulation conceptually clear and allows embedding the effect of the direction of damage in the constitutive law
automatically. After the microplane stress vectors $\boldsymbol\sigma_P$ are computed, the macroscopic stress tensor is defined in a variational
sense through the \emph{principle of virtual work}:
 \begin{equation}\label{microvirtualwork}
\boldsymbol\sigma=\frac 3 {2\pi} \int_{\Omega}\boldsymbol{\mathcal{P}}^T\boldsymbol\sigma_{P}\text{d}\Omega
 \end{equation}
where $\Omega$ is the surface of a unit sphere representing all the possible microplane orientations.
\subsection{Spectral decomposition of the elastic tensor}
In the microplane formulation, the material anisotropy is addressed by decomposing the stress and strain tensors
into energetically orthogonal modes through the \textit{spectral stiffness
decomposition theorem} \cite{Rychlewski95,TheSok98,TheSok99,TheSok00}. The following sections are intended to provide a brief introduction of the theory.

\subsubsection{Spectral decomposition of the elastic tensor}
The elastic behavior of a general anisotropic material can be expressed in \emph{Kelvin notation} \cite{Thomson78,Elbing94} as:
 \begin{equation}\label{e3}
  \boldsymbol\sigma=\boldsymbol{\mathcal{C}} \boldsymbol\varepsilon
 \end{equation}
where $\boldsymbol\sigma=\left[\sigma_{11}~\sigma_{22}~\sigma_{33}~\sqrt{2}\sigma_{23}~\sqrt{2}\sigma_{13}~\sqrt{2}\sigma_{12} \right]^T$,
$\boldsymbol \varepsilon=\left[\varepsilon_{11}~\varepsilon_{22}~\varepsilon_{33}~\sqrt{2}\varepsilon_{23}~\sqrt{2}\varepsilon_{13}~\sqrt{2}\varepsilon_{12} \right]^T$
are the contracted forms of the stress and strain second-order tensors and $\boldsymbol{\mathcal{C}}$ represents the contracted form of the fourth-order elastic tensor.
The indices refer to Cartesian coordinates $x_i$ ($i=1,2,3$) as defined in Figures (\ref{f1}a,b). It is worth mentioning here that the factor $\sqrt{2}$ assures that both the stiffness tensor and its
column matrix have the same norm, given by the sum of the squares of their elements.

According to the spectral decomposition theorem \cite{Rychlewski95,TheSok98,TheSok99,TheSok00}, the stiffness matrix $\boldsymbol{\mathcal{C}}$ can be decomposed as follows:
 \begin{equation}\label{e4}
 \boldsymbol{\mathcal{C}}=\sum_I \lambda^{(I)}\boldsymbol{\mathcal{C}}^{(I)}
 \end{equation}
where $I=1,2...6$,  $\lambda^{(I)}$ are the eigenvalues of the stiffness matrix and $\boldsymbol{\mathcal{C}}^{(I)}=\sum_n \boldsymbol\phi_{In}\boldsymbol\phi_{In}^T$ are a set of
second-order tensors constructed from the elastic eigenvectors $\boldsymbol\phi_I$. The $I$-th eigenvector $\boldsymbol\phi_I$ has multiplicity $n$ and is normalized such
that $\boldsymbol\phi_I^T\boldsymbol{\mathcal{C}}^{(I)}\boldsymbol\phi_I=\lambda^{(I)}$.

In general, the elastic eigenvalues and eigenmatrices can be found by solving the following eigenvalue problem:
 \begin{equation}\label{e5}
 \left[\boldsymbol{\mathcal{C}}-\lambda^{(I)}\boldsymbol{\mathcal{I}}\right]\boldsymbol{v}=0
 \end{equation}
where, it should be noted, the number of independent elastic eigenvalues is strictly related the degree of anisotropy of the material.
In general anisotropy, the solution of Eq. (\ref{e5}) provides $6$ independent eigenvalues which are functions of the
$21$ independent elastic constants of the material.

It is worth mentioning that the eigenmatrices $\boldsymbol{\mathcal{C}}^{(I)}$ are:
\begin{enumerate}
  \ii  \emph{partition of unity}, i.e. $\sum_I\boldsymbol{\mathcal{C}}^{(I)}=\boldsymbol{\mathcal{I}}$;
  \ii  \emph{orthogonal}, i.e. $\boldsymbol{\mathcal{C}}^{(I)}\boldsymbol{\mathcal{C}}^{(J)}=0$ if $I\ne J$;
  \ii  \emph{idempotent}, i.e. $\boldsymbol{\mathit{C}}^{(I)}\boldsymbol{\mathcal{C}}^{(I)}=\boldsymbol{\mathcal{C}}^{(I)}$.
\end{enumerate}
Moreover, it should be remarked that the property 2) above leads to energetic orthogonality: $ \delta\mathcal{W}_{IJ}=\boldsymbol\sigma_I^T\delta\boldsymbol\varepsilon_J=\lambda^{(I)}\boldsymbol\varepsilon_I^T\delta\boldsymbol\varepsilon_J=\lambda^{(I)}\boldsymbol\varepsilon^T \bold{C}^{(I)} \bold{C}^{(J)}\delta\boldsymbol\varepsilon=0$.

An important characteristic of the elastic eigenmatrices $\boldsymbol{\mathcal{C}}^{(I)}$ is that they provide
a way to decompose the stress and strain tensors into energetically orthogonal modes. These are called here
\emph{eigenstresses} and \emph{eigenstrains} and are defined as:
 \begin{equation}\label{e6}
 \boldsymbol\sigma_I=\boldsymbol{\mathcal{C}}^{(I)} \boldsymbol\sigma \quad \mbox{and} \quad  \boldsymbol\varepsilon_I=\boldsymbol{C}^{(I)} \boldsymbol\varepsilon
 \end{equation}
It is easy to show that $\boldsymbol\sigma=\sum_I\boldsymbol\sigma_I$ and $\boldsymbol\varepsilon=\sum_I\boldsymbol\varepsilon_I$ whereas the relation
between eigenstresses and eigenstrains can be found introducing the related elastic eigenvalues: $\boldsymbol\sigma_I=\lambda^{(I)}\boldsymbol\varepsilon_I$.

A similar analysis can be applied to the compliance elastic matrix $\boldsymbol{\mathcal{S}}=\boldsymbol{\mathcal{C}}^{-1}$.
In this case, the relation between the compliance matrix and its eigenmatrices reads:
 \begin{equation}\label{e7}
 \boldsymbol{\mathcal{S}}=\sum_I \left(\lambda^{(I)}\right)^{-1}\boldsymbol{\mathcal{S}}^{(I)}
 \end{equation}
where it could be easily shown that the compliance eigemenmatrices coincide with the ones derived from the stiffness tensor, $\boldsymbol{\mathcal{S}}^{(I)}\equiv \boldsymbol{\mathcal{C}}^{(I)}$.

\subsubsection{Microplane model with spectral decomposition}
In this section, the decomposition of the strain tensor into energetically orthogonal eigenstrains through the spectral decomposition of the
elastic tensor is adopted to extend the microplane approach to general anisotropic materials.
By the spectral decomposition of the strain tensor and a separate projection of each eigenstrain, each microplane vector can be
decomposed into microplane eigenstrain vectors as:
 \begin{equation}\label{microeigenstrain}
 \boldsymbol\varepsilon_P=\sum_I^N\boldsymbol\varepsilon_{P}^{(I)}\qquad\mbox{where}~
 \begin{cases}
 &\boldsymbol\varepsilon_{P}^{(I)}=\boldsymbol{\mathcal{P}}\boldsymbol\varepsilon^{(I)}=\boldsymbol{\mathcal{P}}^{(I)}\boldsymbol\varepsilon
 \\
 &\boldsymbol{\mathcal{P}}^{(I)}=\boldsymbol{\mathcal{P}}\boldsymbol{\mathcal{C}}^{(I)}
 \end{cases}
 \end{equation}
where $N=$ number of independent eigenmodes.

The main advantage is that now, different constitutive laws describing the material behavior at the microplane level can be
related to each eigenmode. As it will be clear in the following sections, this not only allows the description of the material
anisotropy but also to address the different damaging mechanisms related to different loading conditions.
Accordingly, from the microplane eigenstrains, the microplane eigenstresses $\boldsymbol\sigma_{P}^{(I)}$ can be defined through specific
constitutive laws: $\boldsymbol\sigma_{P}^{(I)}=f\left(\boldsymbol\varepsilon_{P1},\boldsymbol\varepsilon_{P2}...\right)\boldsymbol\varepsilon_{P}^{(I)}$.
Then, substituting the microplane eigestresses into Eq.(\ref{microvirtualwork}) and recalling Eq. (\ref{e6}), the macroscopic stress tensor
can be computed in a variational sense through the principle of virtual work \cite{CusBegBaz08}:
 \begin{equation}\label{eigenvirtualwork}
\boldsymbol\sigma=\frac 3 {2\pi} \int_{\Omega}\boldsymbol{\mathcal{P}}^T\sum_I^N\boldsymbol\sigma_{P}^{(I)}\text{d}\Omega
 \end{equation}
where $\Omega$ is the surface of a unit sphere representing all the possible microplane orientations.
It is worth mentioning here that Eq. (\ref{eigenvirtualwork}) represents a weak variational constraint. In general, the projection
of the stress tensor does not coincide with the microplane eigenstress, i.e. $\boldsymbol\sigma_{P}^{(I)}\ne \boldsymbol{\mathcal{P}}^{(I)}\boldsymbol\sigma$. Such a coincidence, called
\emph{double constraint}, holds in the elastic regime if and only if the microplane eigenstress vector is proportional to microplane
eigenstrain vector through the related eigenvalue, as shown by Cusatis \emph{et al.} \cite{CusBegBaz08}: $\boldsymbol\sigma_{P}^{(I)}=\lambda^{(I)}\boldsymbol\varepsilon_{P}^{(I)}$.

\subsection{Analysis of microplane eigenstrain modes and physical interpretation}\label{interpretation}
It is useful to study the distribution of the normal components of each
microplane eigenstrain
on the microplane sphere to give a physical interpretation. This is key to define physically-based constitutive
laws capable of addressing both material and damage-induced anisotropy.

\subsubsection{Strain decomposition for a twill 2$\times$2 composite}
In order to exemplify the application of the spectral stiffness decomposition theorem to the microplane model,
let us consider the case of a textile composite reinforced by a 2$\times$2 twill fabric made of carbon fibers. The elastic properties of the material can
be found in Table (\ref{T2}) while the Representative Unit Cell (RUC) for this material is represented in Figure (\ref{f1}a) together with the local coordinate system used to define the elastic tensor of the material. As can be noted, planes 1-2, 1-3 and 2-3 represent 3 planes of material symmetry so that the material can be considered as orthotropic \cite{MisKar99}.

For an orthotropic material, the number of independent elastic constants is 9. However, for the case of a twill 2$\times$2, the in-plane elastic moduli as well as the out-of-plane Poisson ratios and shear moduli are usually almost the same (i.e. $E_3\approx E_2=E$, $\nu_{13}\approx \nu_{12}=\nu^'$ $G_{13}\approx G_{12}=G^'$ respectively). Accordingly, the elastic compliance tensor, $\boldsymbol{\mathcal{S}}$, can be expressed as a function of only 6 elastic constants as:
\begin{equation}\label{compliance}
 \boldsymbol{\mathcal{S}}= \begin{bmatrix} 1/E^' & -\nu^'/E & -\nu^'/E & 0 & 0 & 0 \\ -\nu^'/E &
  1/E & -\nu/E & 0 & 0 & 0 \\ -\nu^'/E & -\nu/E & 1/E & 0 & 0 & 0   \\ 0 & 0 & 0 & 1/2G & 0 & 0 \\  0 & 0 & 0 & 0 & 1/2G' & 0 \\ 0 & 0 & 0 &    0 & 0 & 1/2G' \end{bmatrix}
\end{equation}
where $\nu=\nu_{23}=\nu_{32}=$ in-plane Poisson ratio and $G=G_{23}=$ in-plane shear modulus. The elastic stiffness tensor can be computed as $\boldsymbol{\mathcal{C}}=\boldsymbol{\mathcal{S}}^{-1}$.

Thanks to the material symmetry, the solution of the elastic eigenvalue problem in Eq. (\ref{e5}) can be carried out analytically giving the following 5 independent elastic eigenvalues:
\begin{align}
\label{e8}
\begin{split}
&\left(\lambda^{(1)}\right)^{-1}=\frac{1}{2E^'}+\frac{1-\nu}{2E}-\sqrt{\left(\frac{1-\nu}{2E}-\frac{1}{2E^'}\right)^2+2\left(\frac{\nu^'}{E^'}\right)^2}
\\
&\left(\lambda^{(2)}\right)^{-1}=\frac{1+\nu}{E}
\\
&\left(\lambda^{(3)}\right)^{-1}=\frac{1}{2E^'}+\frac{1-\nu}{2E^'}+\sqrt{\left(\frac{1-\nu}{2E}-\frac{1}{2E^'}\right)^2+2\left(\frac{\nu^'}{E^'}\right)^2}
\\
&\left(\lambda^{(4)}\right)^{-1}=\frac{1}{2G^'}
\\
&\left(\lambda^{(5)}\right)^{-1}=\frac{1}{2G}
\end{split}
\end{align}
The related eigematrices, constructed from the normalized eigenvectors, are:
\begin{align}
\label{e9}
\begin{split}
&C_{ij}^{(1)}=\frac {\chi^2\delta_{1j}+\chi \left(\delta_{1j}\delta_{2i}+\delta_{2j}\delta_{1i}+\delta_{1j}\delta_{3i}+\delta_{3j}\delta_{1i}\right)+\delta_{2j}\left(\delta_{i2}+\delta_{3i}\right)+\delta_{3j}\left(\delta_{i3}+\delta_{2i}\right)}{{2+\chi^2}}
\\
&C_{ij}^{(2)}=\frac{(\delta_{2j}-\delta_{3j})(\delta_{i2}-\delta_{i3})}2
\\
&C_{ij}^{(3)}=\frac {\xi^2\delta_{1j}+\xi \left(\delta_{1j}\delta_{2i}+\delta_{2j}\delta_{1i}+\delta_{1j}\delta_{3i}+\delta_{3j}\delta_{1i}\right)+\delta_{2j}\left(\delta_{i2}+\delta_{3i}\right)+\delta_{3j}\left(\delta_{i3}+\delta_{2i}\right)}{{2+\xi^2}}
\\
&C_{ij}^{(4)}=\delta_{i5}\delta_{5j}+\delta_{i6}\delta_{6j}
\\
&C_{ij}^{(5)}=\delta_{i4}\delta_{4j}
\end{split}
\end{align}
 where $\delta_{ij}=$ Kronecker delta, $i,j=1,2...6$ and:
 \begin{align}
\label{e10}
\begin{split}
&\chi=\frac{E^'(1-\nu)}{2E\nu^'}-\frac{1}{2\nu^'}+\sqrt{\left[\frac{E^'(1-\nu)}{2E\nu^'}-\frac{1}{2\nu^'}\right]^2+2}
\\
&\xi=\frac{E^'(1-\nu)}{2E\nu^'}-\frac{1}{2\nu^'}-\sqrt{\left[\frac{E^'(1-\nu)}{2E\nu^'}-\frac{1}{2\nu^'}\right]^2+2}
\end{split}
\end{align}
are dimensionless constants which depend on the elastic properties of the material system.

\subsubsection{Microplane eigenstrain distribution}
After the spectral decomposition of the stiffness matrix, the expressions for the normal, $\varepsilon_N^{(I)}$, and total shear $\varepsilon_T^{(I)}=\sqrt{(\varepsilon_M^{(I)})^2+(\varepsilon_L^{(I)})^2}$
components can be easily computed substituting Eqs.(\ref{e2}) and (\ref{e9}a-e) into Eq. (\ref{microeigenstrain}). As can be noted from their expressions, reported in Table (\ref{T1}), the microplane
eigenvector components depend on 1) the the applied macroscopic strains through the functions $\alpha=\left(\varepsilon_2+\varepsilon_3+\varepsilon_1 \chi\right)/\left(2+\chi^2\right)$, $\beta=1/2\left(\varepsilon_3-\varepsilon_2\right)$,
$\gamma=\left(\varepsilon_2+\varepsilon_3+\varepsilon_1 \zeta\right)/\left(2+\zeta^2\right)$, 2) the orientation of the microplane normals and 3) the material
elastic properties through the dimensionless constants $\xi$ and $\chi$ defined in Eqs.(\ref{e10}a,b).

According to Eqs. (\ref{e10}a,b), $\xi=-10.0$ and $\chi=0.2$ respectively for the material under study
and the distribution of normal and shear components can now be analyzed for different applied strains at
the macroscale. Let us now consider the distribution of the normal strain components
on the microplane sphere caused by a macroscale uniaxial strain applied
along the $x_2$-axis or the $x_3$-axis (Fig. \ref{f2}a,b). Strain mode 4 and strain mode 5 are exactly zero because
they do not depend on $\varepsilon_2$ and $\varepsilon_3$ whereas mode 3 is nonzero but negligible compared to modes 1 and 2.
As can be noted from Figure (\ref{f2}), which represents the distribution of the normal components of mode 1 and 2 on the microplane sphere, mode 1 acts as a \textit{volumetric-like} mode, being uniformly distributed along the $2$-$3$ plane. Conversely, mode 2 acts as a \textit{deviatoric-like} mode, loading mainly the microplanes whose normal is close to the direction of application of the macroscale strain. The two modes have almost the same magnitude in the direction of application of the macroscale strain (Figure \ref{f2}a,b).
Since both modes describe the uniaxial behavior of the material system in the plane of the fabric, it is convenient to consider the effect of the two modes summed together: $\boldsymbol\varepsilon_P^{(12)}=\lambda_1/\lambda_2\boldsymbol\varepsilon_P^{(1)}+\boldsymbol\varepsilon_P^{(2)}$. The resulting mode will be referred as \textit{mode 12} in the following. It should be noted that mode 1 is weighted by the ratio between the first and the second eigenvalues of the elastic tensor, which for most of textile composites is within $0.8-1.2$. The reason of this choice will be clear in section \ref{inelastic12}. As Figure \ref{f2} shows, mode 12 describes the in-plane uniaxial behavior of the material in the direction of application of the load. Being the in-plane behavior a ``fiber-dominated" property, this analysis shows that the response of the composite material subjected to strain mode 12 strongly depends on the behavior of the fibers.

A different picture arises if one considers a uniaxial strain at the
macroscale applied in the out-of-plane direction, i.e. $x_1$-axis. Strain mode 4 and strain mode 5 are exactly zero because
they do not depend on $\varepsilon_1$ while mode 12 is nonzero but negligible compared to mode 3. Figure \ref{f3} shows the
distribution of the normal strain component. It can be concluded that mode 3 is mainly related to the behavior of the material system when
subjected to an out-of-plane uniaxial strain at the macroscale, $\varepsilon_1$. The behavior in this direction is expected to be ``matrix-dominated" as the fabric and the matrix mainly act in series coupling in this direction of loading.

As can be expected from the equations describing its microplane strain components (Table \ref{T1}), mode 4 is activated only in the presence of
a macroscopic in-plane shear strain, i.e. $\varepsilon_4$. In this configuration, all the other modes are zero while mode 4 acts mainly on microplanes oriented in direction $\theta=\pi/4$. The in-plane shear behavior is highly dictated by the matrix. Accordingly, as for mode 3, mode 4 can be considered as a ``matrix dominated mode''.
Similar considerations can be drawn for mode 5, this mode being activated in the presence of macroscopic out-of-plane shear strains, i.e. $\varepsilon_5$ and $\varepsilon_6$.
For this case, the distribution of the normal component of the microplane eigenstrain is shown in Figure \ref{f4}.

\subsection{Formulation of microplane constitutive laws}
In this section, the microplane constitutive laws which, for each eigenmode, provide the
relationship between the strain vector and the stress vector at
the microplane level, are presented.
\subsubsection{Elastic behavior}
The elastic behavior is formulated by assuming that normal and
shear eigenstresses on the microplanes are proportional to the corresponding eigenstrains:
  \begin{equation}\label{e11}
\sigma_N^{(I)}=\lambda^{(I)}\varepsilon_N^{(I)},\quad \sigma_M^{(I)}=\lambda^{(I)}\varepsilon_M^{(I)}, \quad \sigma_L^{(I)}=\lambda^{(I)}\varepsilon_L^{(I)}
 \end{equation}
where $\lambda^{(I)}=$ $I$-th elastic eigenvalue. It should be noted that, as proved by Cusatis \emph{et al}. in \cite{CusBegBaz08}, these relations guarantee a double constraint
in the elastic regime, i.e. the projection of
the stress tensor does coincide with the microplane eigenstress: $\boldsymbol\sigma_{P}^{(I)}= \boldsymbol{\mathcal{P}}^{(I)}\boldsymbol\sigma$. This condition is not generally true and it is violated as soon as the material reaches the inelastic regime.
\subsubsection{Inelastic behavior: general constitutive law at microplane level}
The spectral stiffness microplane formulation of the nonlinear and inelastic behavior
aims at representing the main physical mechanisms characterizing the meso-scale failure such as
e.g. \emph{fiber failure}, \emph{fiber-matrix debonding} and \emph{matrix microcracking}.

Let us consider a generic $I$-th eigenmode and, following Cusatis \emph{et al.} \cite{CusBazCed03,CusPelMen11}, define a microplane \emph{effective eigenstrain} as:
  \begin{equation}\label{e12}
\varepsilon^{(I)}=\sqrt{(\varepsilon_{N}^{(I)})^2+(\varepsilon_{T}^{(I)})^2}
 \end{equation}
 where $\varepsilon_{T}^{(I)}=\sqrt{(\varepsilon_{M}^{(I)})^2+(\varepsilon_{L}^{(I)})^2}$= total shear strain component of $I$-th microplane eigenstrain.
 The constitutive law can then be defined by means of an \emph{effective eigenstress}, $\sigma^{(I)}$. The relation between the stress and strain
 microplane components can be found imposing the consistency of the virtual work:
   \begin{equation}\label{e13}
\delta \mathcal{W}_{I}=\sigma^{(I)}\delta \varepsilon^{(I)}=\frac{\sigma^{(I)}}{\varepsilon^{(I)}}\left(\varepsilon_{N}\delta\varepsilon_{N}+ \varepsilon_{M}\delta\varepsilon_{M}+\varepsilon_{L}\delta\varepsilon_{L}\right)^{(I)}= \left(\sigma_{N}\delta\varepsilon_{N}\right)^{(I)}+ \left(\sigma_{M}\delta\varepsilon_{M}\right)^{(I)}+\left(\sigma_{L}\delta\varepsilon_{L}\right)^{(I)}
 \end{equation}
where, it should be noted, thanks to the definition in Eq. (\ref{e12}) and to Eq. (\ref{e13}), $\delta \mathcal{W}_{I}\ge 0$ for all the microplanes. This means that energy dissipation on each microplane is non-negative, a \textit{sufficient} condition to fulfill the second law of thermodynamics.

By means of Eq. (\ref{e13}), the relationship between normal and shear stresses versus normal and
shear strains can be formulated through damage-type constitutive equations:
   \begin{equation}\label{e14}
\sigma_{N}^{(I)}=\left(\sigma\frac{\varepsilon_{N}}{\varepsilon}\right)^{(I)}, \quad \sigma_{M}^{(I)}=\left(\sigma\frac{\varepsilon_{M}}{\varepsilon}\right)^{(I)}, \quad \sigma_{L}^{(I)}=\left(\sigma\frac{\varepsilon_{L}}{\varepsilon}\right)^{(I)}
 \end{equation}
It is worth pointing out that, substituting Eqs. (\ref{e14}) into (\ref{e12}) and rearranging, leads to the following definition of the $I$-th effective stress: $\sigma^{(I)}=\sqrt{(\sigma_{N}^{(I)})^2+(\sigma_{T}^{(I)})^2}$ where $\sigma_{T}^{(I)}=\sqrt{(\sigma_{M}^{(I)})^2+(\sigma_{L}^{(I)})^2}$= total shear stress component.

The effective stress $\sigma^{(I)}$ is assumed to be incrementally elastic, i.e. $\dot{\sigma}^{(I)}=\lambda^{(I)}\dot{\varepsilon}^{(I)}$
and it is formulated such that $ 0\le\sigma^{(I)}\le \sigma_{bi}^{(I)}(\varepsilon^{(12)},\varepsilon^{(3)}...,\theta,\varphi)$ where
$\sigma_{bi}^{(I)}(\varepsilon^{(12)},\varepsilon^{(3)}...,\theta,\varphi)$ with subscript $i=$t for tension and $i=$c for compression is a limiting boundary enforced through a vertical (at constant strain) return algorithm.
It is worth mentioning here that, in general, $\sigma_{bi}^{(I)}$ is a function of the microplane orientation and of the equivalent strains pertaining to other modes.
This allows to inherently embed in the formulation the effects of damage anisotropy and the interaction between damaging mechanisms.

As clarified in section \ref{interaction}, each mode can be ascribed to a particular micromechanical constituent and a type of deformation. This make them particulary suitable for the description of the diverse damage mechanisms of the material through eigenmode-dedicated boundaries, the description of which being the subject of the following sections.

\subsubsection{Inelastic behavior: mode 12}\label{inelastic12}
As discussed in section \ref{interpretation}, since both modes 1 and 2 describe the fiber behavior in the
presence of a macroscopic uniaxial in-plane strain, it is convenient to combine them together.
Accordingly, the formulation of the effective strain and stress has to be changed to
guarantee work consistency.

Let us define mode 12 effective microplane eigenstrain as:
  \begin{equation}\label{e15}
\varepsilon^{(12)}=\sqrt{\frac{\lambda^{(1)}}{\lambda^{(2)}}\left[(\varepsilon_{N}^{(1)})^2+(\varepsilon_{T}^{(1)})^2\right]+(\varepsilon_{N}^{(2)})^2+(\varepsilon_{T}^{(2)})^2}
 \end{equation}
 After introducing the equivalent stress, $\sigma^{(12)}$, the virtual work can be calculated as done in Eq. (\ref{e13}):
   \begin{equation}\label{e16}
\delta \mathcal{W}_{12}=\frac{\sigma^{(12)}}{\varepsilon^{(12)}}\left[\frac{\lambda^{(1)}}{\lambda^{(2)}}\left(\varepsilon_{N}^{(1)}\delta\varepsilon_{N}^{(1)}+ \varepsilon_{M}^{(1)}\delta\varepsilon_{M}^{(1)}+\varepsilon_{L}^{(1)}\delta\varepsilon_{L}^{(1)}\right)+\varepsilon_{N}^{(2)}\delta\varepsilon_{N}^{(2)}+ \varepsilon_{M}^{(2)}\delta\varepsilon_{M}^{(2)}+\varepsilon_{L}^{(2)}\delta\varepsilon_{L}^{(2)}\right]
 \end{equation}
 Now, recalling that $\delta \mathcal{W}_{12}= \sum_{I=1}^2\sigma_{N}^{(I)}\delta\varepsilon_{N}^{(I)}+ \sigma_{M}^{(I)}\delta\varepsilon_{M}^{(I)}+\sigma_{L}^{(I)}\delta\varepsilon_{L}^{(I)}$,
the relationship between normal and shear stresses versus normal and shear strains can be formulated through the following damage-type constitutive equations:
\begin{equation}\label{e17}
\begin{split}
&\sigma_{N}^{(12)}=\frac{\sigma^{(12)}}{\varepsilon^{(12)}}\left(\frac{\lambda^{(1)}}{\lambda^{(2)}}\varepsilon_{N}^{(1)}+\varepsilon_{N}^{(2)}\right),~~\varepsilon_{M}^{(12)}=\frac{\sigma^{(12)}}{\varepsilon^{(12)}}\left(\frac{\lambda^{(1)}}{\lambda^{(2)}}\varepsilon_{M}^{(1)}+\varepsilon_{M}^{(2)}\right),~~\sigma_{L}^{(12)}=\frac{\sigma^{(12)}}{\varepsilon^{(12)}}\left(\frac{\lambda^{(1)}}{\lambda^{(2)}}\varepsilon_{L}^{(1)}+\varepsilon_{L}^{(2)}\right)
\end{split}
 \end{equation}
Again, combining Eqs. (\ref{e17}) with (\ref{e15}) one gets the new definition of the effective stress: $\sigma^{(12)}=\sqrt{\lambda^{(2)}/\lambda^{(1)}[(\sigma_{N}^{(1)})^2+(\sigma_{T}^{(1)})^2]+(\sigma_{N}^{(2)})^2+(\sigma_{T}^{(2)})^2}$.  The aforementioned definition of the effective strain and stress guarantees the correct description of the elastic behavior. As a matter of fact, assuming the effective stress $\sigma^{(12)}$ to be incrementally elastic, i.e. $\dot{\sigma}^{(12)}=\lambda^{(2)}\dot{\varepsilon}^{(12)}$, one obtains again Eqs. (\ref{e11}).

The strain dependent limiting boundary, $\sigma_{bi}^{(12)}(\varepsilon^{(12)},\varepsilon^{(3)},...,\theta,\varphi)$, is expressed as follows:
    \begin{equation}\label{e18}
    \begin{cases}
\sigma_{bt}^{(12)}=s^{(12)}\left(\theta,\varphi,\varepsilon_{\max}^{(5)}\right)\exp\left[-\left(\frac{\left\langle \varepsilon_{\max}^{(12-t)}-\varepsilon_{0t}^{(12)}\right\rangle}{k_{bt}^{(12)}}\right)^{a_{t12}} \right] \qquad \mbox{for}~\frac{\lambda^{(1)}}{\lambda^{(2)}}\varepsilon_{N}^{(1)}+\varepsilon_{N}^{(2)}\ge 0
\\
\sigma_{bc}^{(12)}=c^{(12)}\left(\theta,\varphi,\varepsilon_{\max}^{(5)}\right)\exp\left[-\left(\frac{\left\langle \varepsilon_{\max}^{(12-c)}-\varepsilon_{0c}^{(12)}\right\rangle}{k_{bc}^{(12)}}\right)^{a_{c12}} \right] \qquad \mbox{for}~\frac{\lambda^{(1)}}{\lambda^{(2)}}\varepsilon_{N}^{(1)}+\varepsilon_{N}^{(2)} < 0
\end{cases}
 \end{equation}
 where the brackets $\langle \bullet \rangle$ are used in Macaulay sense: $\langle x \rangle = \max (x,0)$. The functions $s^{(12)}(\theta,\varphi,\varepsilon_{\max}^{(4)})=s_0^{(12)}(\varepsilon_{\max}^{(4)})$ and $c^{(12)}(\theta,\varphi,\varepsilon_{\max}^{(4)})=c_0^{(12)}(\varepsilon_{\max}^{(4)})$ represent mode $12$ microplane tensile and compressive strength respectively which, in general, depend on the microplane orientation. However, for the material under study and for most of symmetric textile composites in which the behavior in the warp direction is similar to the weft direction, they can be both assumed constant.

 The microplane strengths depend, in general, on the maximum effective mode 4 strain, $\varepsilon_{\max}^{(4)}$. The interaction between the different modes will be clarified in section \ref{interaction}.

As can be seen in Figure (\ref{f6}a), the boundary $\sigma_{bi}^{12}$ evolves exponentially as a function of the maximum effective strain, which is a history-dependent variable defined as
$\varepsilon_{\max}^{(12-i)}(t)=\max_{\tau\le t}[\varepsilon^{(12)}(t)]$ with $i=$t for tension and $i=$c for compression. The exponential decay of the boundary $\sigma_{bi}^{12}$ starts when the maximum effective strain reaches its elastic limit $\varepsilon_{0t}^{(12)}(\theta,\varphi,\varepsilon_{\max}^{(5)})=s^{(12)}/\lambda^{(12)}$ and $\varepsilon_{0c}^{(12)}(\theta,\varphi,\varepsilon_{\max}^{(5)})=c^{(12)}/\lambda^{(12)}$ for tension and compression respectively. The decay rate is governed by the post-peak slope (Figure \ref{f6}a) which, as can be derived from Eq. (\ref{e18}), reads:
   \begin{equation}\label{e19}
\begin{cases}
&H_t\left(\varepsilon_{\max}^{(12)}\right)=\frac{\sigma_{bt}^{12}a_{t12}}{k_{bt}^{(12)}}\left(\frac{\varepsilon_{\max}^{(12-t)}-s^{(12)}/\lambda^{(12)}}{k_{bt}^{(12)}}\right)^{a_{t12}-1}\quad\mbox{in tension}
\\
&H_c\left(\varepsilon_{\max}^{(12)}\right)=\frac{\sigma_{bt}^{12}a_{c12}}{k_{bc}^{(12)}}\left(\frac{\varepsilon_{\max}^{(12-c)}-c^{(12)}/\lambda^{(12)}}{k_{bc}^{(12)}}\right)^{a_{c12}-1}\quad\mbox{in compression}
\end{cases}
 \end{equation}
For $a_{t12},a_{c12} \le 1$ the initial softening modulus goes to infinity, i.e. the curve has an initial vertical slope. This seems to match perfectly the in-plane behavior of composites which behave elastically up to the peak in stress before softening.
In general, the total number of required parameters to the describe mode $12$ in tension and compression is $6$. However, in absence of specific experimental data, the exponents $a_{t12},a_{c12}$ can be set equal, as done in the present work (see Table \ref{T3}).

\subsubsection{Inelastic behavior: mode 3}\label{inelastic3}
As shown in the previous sections, the 3rd mode can be related to the normal deformation in the out-of-plane direction. The strain dependent boundary can be expressed by the following equations:
    \begin{equation}\label{e20}
    \begin{cases}
\sigma_{bt}^{(3)}=s^{(3)}\left(\theta,\varphi\right)\left\langle 1- \frac{\left\langle \varepsilon_{\max}^{(3-t)}-k_{at}^{(3)}\right\rangle}{k_{bt}^{(3)}} \right\rangle \qquad \mbox{for}~\varepsilon_{N3}\ge 0
\\
\sigma_{bc}^{(3)}=c^{(3)}\left(\theta,\varphi\right)\left\langle 1- \frac{\left\langle \varepsilon_{\max}^{(3-c)}-k_{ac}^{(3)}\right\rangle}{k_{bc}^{(3)}} \right\rangle \qquad \mbox{for}~\varepsilon_{N3} < 0
\end{cases}
 \end{equation}
where $s^{(3)}\left(\theta,\varphi\right)=s_0^{(3)}=$ mode $3$ microplane tensile strength, $c^{(3)}\left(\theta,\varphi\right)=c_0^{(3)}=$ mode $3$ microplane compressive strength. These functions are both assumed to be independent of the microplane orientation. As can be seen in Figure (\ref{f6}b), Eqs. (\ref{e20}a,b) result into a bilinear boundary in tension and compression. The parameters $k_{at}^{(3)}$ and $k_{ac}^{(3)}$ represent the value of the effective strain at the
beginning of the linear softening region whereas the parameters $k_{bt}^{(3)}$ and $k_{bc}^{(3)}$ are defined such that
$k_{bt}^{(3)}+k_{at}^{(3)}$ and $k_{bc}^{(3)}+k_{ac}^{(3)}$ represent the value of the effective strain when $\sigma_{bi}^{(3)}=0$ in tension and compression respectively.
The total number of required parameters to describe mode $3$ in tension and compression is $6$. However, in most of applications of composite laminates, the out-of-plane stress is often negligible. Accordingly, the material is likely to behave elastically in this direction so that an accurate characterization of the parameters related to mode 3 is not required. In the present work, in the absence of direct experimental data, mode 3 parameters are estimated from \cite{SodHinKad98,SodHinKad02} assuming that the out-of-plane behavior is ``matrix-dominated".

\subsubsection{Inelastic behavior: mode 4}
The 4th mode is related to in-plane shear deformation. In this type of deformation, the material is subjected to extensive microcracking before reaching the peak in stress which produces a remarkable non-linear behavior. In order to capture this phenomenon, the strain dependent boundary is expressed by the following equations:
     \begin{equation}\label{e21}
\sigma_{bt}^{(4)} =\left(s^{(4)}\right)^{1-p}\left(\lambda^{(4)}\right)^p \begin{cases} \left(\varepsilon_{\max}^{(4-t)}\right)^p, &\qquad  \mbox{if }  \varepsilon_{\max}^{(4-t)}\le s^{(4)}/\lambda^{(4)}\\ \left(k_{at}^{(4)}\right)^p \left\langle 1- \frac{\left\langle \varepsilon_{\max}^{(4-t)}-k_{at}^{(4)}\right\rangle}{k_{bt}^{(4)}} \right\rangle, & \qquad \mbox{if } \varepsilon_{\max}^{(4-t)}> s^{(4)}/\lambda^{(4)} \end{cases}
 \end{equation}
for $\varepsilon_N^{(4)}\ge 0$ and
     \begin{equation}\label{e22}
\sigma_{bc}^{(4)} =\left(c^{(4)}\right)^{1-p}\left(\lambda^{(4)}\right)^p \begin{cases} \left(\varepsilon_{\max}^{(4-c)}\right)^p, &\qquad  \mbox{if }  \varepsilon_{\max}^{(4-c)}\le c^{(4)}/\lambda^{(4)}\\ \left(k_{ac}^{(4)}\right)^p \left\langle 1- \frac{\left\langle \varepsilon_{\max}^{(4-c)}-k_{ac}^{(4)}\right\rangle}{k_{bc}^{(4)}} \right\rangle, & \qquad \mbox{if } \varepsilon_{\max}^{(4-c)}> s^{(4)}/\lambda^{(4)} \end{cases}
 \end{equation}
for $\varepsilon_N^{(4)}< 0$.
The functions $s^{(4)}\left(\theta,\varphi\right)=s_0^{(4)}$ and $c^{(4)}\left(\theta,\varphi\right)=c_0^{(4)}$ represent mode $4$ microplane tensile and compressive stresses at which non-linearity starts and they are both assumed to be independent of the microplane orientation. $p$ is a parameter controlling the slope of the curve in the pre-peak region and it assumed to be the same in tension and compression. The parameters $k_{at}^{(4)}$ and $k_{ac}^{(4)}$ represent the value of the effective strain at the
beginning of the linear softening region whereas the parameters $k_{bt}^{(4)}$ and $k_{bc}^{(4)}$ are defined such that
$k_{bt}^{(4)}+k_{at}^{(4)}$ and $k_{bc}^{(4)}+k_{ac}^{(4)}$ represent the value of the effective strain when $\sigma_{bi}^{(4)}=0$ in tension and compression respectively. A typical boundary resulting from Eqs. (\ref{e21}) and (\ref{e22}) can be seen in Figure (\ref{f6}c).
The total number of required parameters to describe mode $4$ in tension and compression is $7$. In cases in which the macroscopic in-plane shear behavior is independent of the sign of the deformation (which is not always true for textile composites), the same parameters at microplane level can be used in tension and compression. In such a case, considered in the present work, the number of required parameters is 4 as reported in Table \ref{T3}.

\subsubsection{Inelastic behavior: mode 5}\label{inelastic5}
Mode 5 describes the shear deformation in the out-of-plane direction. In the absence of direct experimental data, it is assumed here that the constitutive behavior
at microplane level resembles the one related to mode 4. It is worth noting that the effects of these assumption are considered to be negligible
since, in practical situations, out-of-plane shear deformation will always been within the elastic regime. In facts, inter-laminar cracks
would emerge far before than reaching any nonlinear behavior in the material.
\subsubsection{Interaction between eigenmodes}\label{interaction}
After defining the inelastic boundaries, the interaction between modes needs to be clarified to capture the complex multi-axial behavior of textile composites.

The definition of mode 12 effective strain accounts already for the interaction  between the \emph{volumetric} and the \emph{deviatoric} in-plane components at microplane level. As a matter of facts, the stress peak for mode 12 is reached when $\varepsilon^{(12)}=\varepsilon_{0i}^{(12)}$ with the subscript $i=t$ for tension and $i=c$ for compression. This corresponds to the following microplane strain multi-axial criterion:
\begin{equation}\label{multiaxial12}
\frac{\lambda^{(1)}}{\lambda^{(2)}}\left(\frac{\varepsilon^{(1-i)}_{\max}}{\varepsilon_{0i}^{(12)}}\right)^2+\left(\frac{\varepsilon^{(2-i)}_{\max}}{\varepsilon_{0i}^{(12)}}\right)^2=1
\end{equation}
where the definition of effective strain reported in Eq. (\ref{e11}) has been used for mode 1 and 2. As soon as Eq. (\ref{multiaxial12}) is satisfied, strain softening starts to occur.

The interaction between mode 12 and mode 4 can be described in a similar way:
\begin{equation}\label{multiaxial125}
\left(\frac{\varepsilon^{(12-i)}_{\max}}{\varepsilon_{0i}^{(12)}}\right)^2+\left(\frac{\varepsilon^{(4-i)}_{\max}}{k_{ai}^{(4)}}\right)^2=1
\end{equation}
where again $i=t$ for tension and $i=c$ for compression. As soon as Eq. (\ref{multiaxial125}) is satisfied, the new strains at peak are updated and strain softening starts to occur. This corresponds to a vertical scaling of the softening boundary of a factor of $\sqrt{1-(\varepsilon^{(4-i)*}_{\max}/k_{ai}^{(4)})^2}$ for mode 12 and $\sqrt{1-(\varepsilon^{(12-i)*}_{\max}/\varepsilon_{0i}^{(12)})^2}$ for mode 4. $\varepsilon^{(4-i)*}_{\max}$ and $\varepsilon^{(12-i)*}_{\max}$ represent the values of the effective strain for mode 4 and 12 respectively as soon as Eq. (\ref{multiaxial125}) is fulfilled.

Considering the deformations associated to mode 12 and mode 4, it is easy to understand that the criterion in Eq. (\ref{multiaxial125}) defines the mechanical behavior of the material subjected to simultaneous normal and shear in-plane deformations.

Similar interaction laws for the other modes could also be defined. However, in the absence of experimental data for validation, only the interaction between modes 1, 2 and 4 has been considered in the present work.
\subsubsection{General unloading-reloading rule at microplane level}
The inelastic formulation needs to be completed by
unloading-reloading rules to simulate cycling loadings and unloading of the material next to a damaged zone undergoing softening.
Figure (\ref{f6}d) illustrates the unloading-reloading rule adopted in this work in
terms of effective stress versus effective strain.

Let us
assume that unloading occurs after the effective strain increased
continuously from zero to a certain value, $\varepsilon_{max}^{(I-i)}$ where index $i=$t for tension and $i=$c for compression. The effective stress
decreases elastically until it reaches a zero value and remains constant at zero for further decreases of the effective strain. During
reloading, the effective stress remains zero until the effective strain
reaches the reloading strain limit, $\varepsilon_{tr}^{(I-i)}$, and, beyond this point, the
behavior is incrementally elastic. The reloading strain limit is defined as $\varepsilon_{tr}^{(I-i)}=k_{hi}^{(I)}(\varepsilon_{\max}^{(I-i)}-\sigma_{bi}^{(I)}/\lambda^{(I)})$ where $k_{hi}^{(I)}$ is assumed to be a material parameter which
governs the size of hysteresis cycles and, consequently, the amount of energy that the material can dissipate during cycling loading. For
$k_{hi}^{(I)}= 1$, the dissipated energy density is zero, whereas for
$k_{hi}^{(I)}= 0$, it is a maximum and equal to $\sigma_{bi}^{(I)}\varepsilon_{max}^{(I-i)}$. In the present contribution, in the absence of specific experimental data, the dissipated energy density was set equal to zero for all the eigenmodes.

\section{Materials and tests for calibration and validation}
\subsection{Materials}
 In order to calibrate and validate the model, experiments were conducted on twill 2$\times$2 woven composite specimens manufactured by compression molding. A DGEBA-based epoxy resin was chosen as polymer matrix whereas the reinforcement was provided by a twill 2$\times$2 fabric made of carbon fibers. A constant thickness of approximately 1.9 mm, corresponding to 8 laminae,  was used for all the tests. Two different lay-ups, namely a [$0^{\circ}$]$_8$ and a Quasi-Isotropic (Q.I.) $[0^{\circ}/45^{\circ}/-45^{\circ}/90^{\circ}$]$_{s}$, were tested in order to provide sufficient data for calibration and validation.

\subsection{Uniaxial tests}
Uniaxial tests on coupons were conducted to obtain the elastic properties and strength of the composite. In order to provide sufficient data in tension and compression and to provide information regarding the shear behavior of the material, the following test configurations were adopted:
\begin{enumerate}
\ii \textbf{$0^{\circ}$} configuration (conducted on both lay-ups): where the loading axis is aligned with direction 3 of the RUC (see Fig. \ref{f1}), to yield axial moduli and strength of the composite lamina, both under tension and compression;
\ii \textbf{$45^{\circ}$} configuration (conducted for only the [$0^{\circ}$]$_8$, under tension), where the loading direction is at an angle of $45^{\circ}$ with direction 3 (see Figure \ref{f1}), to yield the shear modulus and strength of the lamina.
\end{enumerate}

The uni-axial tension tests on the [$0^{\circ}$]$_8$ lay-up were used for calibration of the model whereas, those on the QI layup were used for validation (i.e. no parameter of the model was changed to match the experimental results). All the tests were conducted in accordance with ASTM standards \cite{ASTMD3039,ASTMD3410}.

\subsection{Size effect tests}
Following \B~\emph{et al.} \cite{BazDanLi96,BazPla98}, size effect tests were conducted on single-edge-notched tension (SENT) specimens, using the [$0^{\circ}$]$_8$ lay-up. These tests represent an indirect way to characterize the initial inta-laminar fracture energy of the material, $G_f$, which is usually hard to measure directly through uniaxial tests due to snap-back instability \cite{BazCed91, BazDanLi96, CusBegBaz08, CanBazHoo11}.

The tested SENT coupons were of three different sizes, scaled as 1:2:4. The dimensions of the smallest coupon were $L$ =100 mm, $D$ = 20 mm and $T$ =1.9 mm. The length $L$ and width $D$ of the medium and large specimen were scaled accordingly while the thickness $T$ remained constant. Each coupon had on one edge a pre-cut notch of length $a = 0.2D$. More details on these tests will be provided in a forthcoming contribution \cite{SalKirBaz15}.
\subsection{Impact tests on composite crash can}\label{drop}
Finally, to verify the predictive capability of the model, impact tests on composite crash cans were conducted. The crush can consisted of two parts manufactured by compression moulding, namely a hat section tube and a reinforcing plate glued together by toughened epoxy glue along flanges (Figure \ref{f11}c,d). Two different lay-up configurations were studied, namely:
\begin{enumerate}
  \ii  hat section tube: $[0^{\circ}]_{11}$, plate: $[0^{\circ}]_{8}$;
  \ii  hat section tube and plate: $[0^{\circ}/90^{\circ}/45^{\circ}/-45^{\circ}/0^{\circ}/90^{\circ}/0^{\circ}/-45^{\circ}/45^{\circ}/90^{\circ}/0^{\circ}]$;
\end{enumerate}
The composite tubes, accurately fixed at the bottom, were impacted by a flat mass of 74.4 kg at a velocity of 4.6 m/s in a drop tower.

\section{Calibration and validation}
\subsection{Finite element implementation}
The present model was implemented in Abaqus Explicit v6.11 as a user material subroutine VUMAT \cite{Abaqus11} and in the special purpose software MARS \cite{MARS}. The structures studied were meshed using four-node shell elements with reduced integration and hourglass control \cite{FlaBel81,BelLev94}.
\subsubsection{Crack band model for strain localization}\label{crackband}
To ensure objective numerical results in the presence of strain localization, the crack band model proposed by \B~\emph{et al.} \cite{Baz76, BazOh83, JirBau12} is adopted. In this approach, the width of the damage localization band, $w_c$, is considered as a material property. This width is also equal to the mesh size, which was here chosen as 2 mm. A change in the element size requires the scaling of the post-peak response of the material such that the fracture energy remains unchanged. The size $w_c$ should not be confused with the size of the RUC, which is roughly 7.5 mm and merely represents the repeating geometric unit of the material.

Thanks to the crack band model, a characteristic size of the material is inherently embedded in the formulation. This is key to correctly describe the intra-laminar fracturing of the material as will be clear in the following sections.
\subsection{Strength and post-peak fracturing behavior}
For the calibration of the model, it is now possible to take advantage of the physical interpretation given to each eigenmode in section \ref{interpretation}. It is known that the strength and fracture behavior in the $0^{\circ}$ configuration is governed by mainly the properties of the fabric, and, in the $45^{\circ}$ configuration, mainly by the matrix properties in shear.

It was shown that mode 12 is the ``fiber-dominated" mode governing the in-plane behavior under normal loading condition whereas mode 4 is the ``matrix-dominated" mode describing the shear behavior. Accordingly, being these modes completely independent, the tests in $0^{\circ}$ configuration can be used to uniquely characterize the constitutive law related to mode 12 whereas the tests in $45^{\circ}$ configuration can be considered for mode 4. Thus, thanks to the clear physical interpretation of each mode, the calibration of constitutive laws becomes straightforward.


In general, the calibration is performed in two steps: (1) Determination of parameters from computations for one material point and (2) verification of structural response from coupon level simulations. The need of a two-step process is explained as follows. As far as the elastic stress-strain behavior and strength are concerned, the response at one material point and of a coupon under a uniform uniaxial load must be the same. However, that is not the case for the post-peak response due to damage localization. At the material point level, a stable post-peak strain softening must exist such that the area under the stress-strain curve is
equal to the fracture energy per unit volume of the crack band. However, just like the experiments, a stable post-peak cannot be obtained in coupon level simulations. To ensure that the model dissipates the correct fracture energy, a material point calibration is required.

\subsubsection{Calibration: mode 12}
As shown in the foregoing sections, mode 12 is strictly related to the in-plane uniaxial behavior of the material. Accordingly, the uniaxial tests on the $[0^{\circ}]_8$ specimens were used for the calibration of the 6 parameters required to describe the inelastic behavior in tension and compression. The parameters $s_0^{(12)}$, $c_0^{(12)}$, $k_{bt}^{(12)}$, $k_{bc}^{(12)}$, $a_{t12}$ and $a_{c12}$ were calibrated in such a way that: (1) the predicted strength matched that from the $[0^{\circ}]_8$ coupon tests, for both tension and compression; (2) the predicted size effect in strength matched the test data (as shown in section \ref{size effect}).

The single material point simulations were conducted with reduced integration on one shell element of size $h_e$ = 2 mm. The calibrated parameters are presented in Table \ref{T3}. It should be noted that, even if the exponents $a_{t12}$ and $a_{c12}$ might have different values, an excellent agreement with experimental data was obtained setting $a_{t12}=a_{c12}=3/4$.

Following the material point calibration, a verification was performed with a coupon level simulation. For the uniaxial tensile test coupon, the dimensions were 100 mm {\sf x} 40 mm {\sf x} 1.9 mm, and for the compressive test they were 100 mm {\sf x} 40 mm {\sf x} 1.9 mm, in accordance with the ASTM standards \cite{ASTMD3039, ASTMD3410}. The finite element models for both coupons were meshed with four-noded shell elements of size 2 mm.

The results for the coupon simulation are shown in Figures (\ref{f7}a,b). As expected, despite a stable post peak at the material point level, a sudden dynamic failure of the coupon is observed at the peak load. This can be seen as the vertical drop of load induced by snap-back instability. Both figures show excellent agreement with experiments under tension as well as compression.

\subsubsection{Calibration: mode 4}
The behavior of the material under off-axis uniaxial loading is highly dominated by the matrix, in particular, by its in-plane shear response. In this configuration, mode 4 is predominant so these tests can be used for its calibration.
The mode 4 parameters $s_0^{(4)}$, $c_0^{(4)}$,$k_{at}^{(4)}$,$k_{ac}^{(4)}$,$k_{bt}^{(4)}$,$k_{bc}^{(4)}$ and $p$ were calibrated at material point level, to match accurately the pre-peak non-linearity and the peak load from the [45]$_8$ coupon. The fracture energy in this configuration was not known. Thus, a conservative value of about 50 N/mm was assumed and the parameters $k_{at}^{(4)}$,$k_{ac}^{(4)}$,$k_{bt}^{(4)}$,$k_{bc}^{(4)}$ regulating the post-peak response were calibrated accordingly.

In the absence of experimental data, the same set of values used in tension was used to calibrate the compressive response. It is worth mentioning, however, that the model does have the capability to describe a different behavior in tension and compression under off-axis loading, a useful feature to describe the shear response of nonsymmetric textile composites.

Following the material point calibration, coupon level simulations were conducted for verification using a 2 mm mesh. The coupon dimensions, in accordance with the ASTM standards, were 100 mm {\sf x} 40 mm {\sf x} 1.9 mm.

Figure (\ref{f7}c) shows the specimen load-displacement curves. It is seen that the shear response is very well matched by the model, including the pre-peak non-linearity caused by constrained matrix microcracking. An instability is observed at the peak load, manifested as a vertical drop of load to zero, which is indicative of a snap-back. The resulting calibrated values are shown in Table \ref{T3}.

It is worth mentioning here again that the results in the $0^{\circ}$ configuration were unaffected by the adjustment of the mode 4 parameters. This means that the model calibration can in general follow a clear sequence, avoiding the complication of an iterative procedure.
\subsubsection{Calibration: mode 3 and mode 5}
As pointed out in sections \ref{inelastic3} and \ref{inelastic5}, mode 3 and mode 5 are not expected to reach their inelastic boundaries as, in most practical cases, the material is considered to be almost in plane stress condition. Accordingly, in the absence of experiments to fully characterize these two modes accurately, their inelastic boundary can be just approximated. The parameters for mode 3 were estimated from \cite{SodHinKad98,SodHinKad02} such that the out-of-plane uniaxial elastic properties and strength match the one of a pure epoxy matrix. Mode 5 parameters were assumed to have the same values as for mode 4.
The calibrated parameters are shown in Table \ref{T4}.

\subsection{Verification and validation}
\subsubsection{Uniaxial tests on Q.I. laminates}
 Subsequently to the parameter calibration, predictions were carried out to validate the model. The tests used for this purpose were uniaxial tests on coupons with a quasi-isotropic lay-up, $[0^{\circ}/45^{\circ}/-45^{\circ}/90^{\circ}$]$_{s}$, for both tension and compression. The coupon dimensions were 100 mm {\sf x} 40 mm {\sf x} 1.9 mm in tension and 100 mm {\sf x} 40 mm {\sf x} 1.9 mm in compression, in accordance with the ASTM standards. Finite element models of the coupons were built and meshed with layered shell elements of size 2 mm, 3 Gauss points being assigned to each layer.

The comparison between experimental and predicted load-displacement curves is plotted in Figures (\ref{f8}a,b). In both cases, the agreement in the initial slope and strength is extremely good. For the tensile tests, the predicted strength is slightly on the lower side. This may be attributed to the fact that the mean strength measured from the $0^{\circ}$ configuration tensile tests, used in calibration, had itself a significant scatter. The prediction of the tensile strength of the QI coupon rests heavily on this calibration. The reason is that the first layer to experience failure is indeed the layer oriented along the loading axis since the strain at failure for this orientation is low. The agreement for the compression case is excellent.

It worth mentioning here again that these results were obtained without changing any of the parameters calibrated before. The excellent agreement with experimental data serves as firm validation of the theory.
\subsubsection{Biaxial failure envelopes}
In order to asses the capability of the model to capture the multi-axial behavior of the composite, the experimental multi-axial failure surfaces reported in \cite{OweGri78,FujAmiLin92} were considered. It is worth mentioning here that the material system studied in these references is a plain weave composite, different from the twill 2$\times$2 composite used to calibrate the model. Accordingly, the comparison between model predictions and the experimental data has to be considered as qualitative only.

Figure (\ref{f10}a) shows the experimental data by Owen and Griffiths \cite{OweGri78} on the failure under normal biaxial loading of a system composed by a polyester matrix reinforced by a glass woven fabric. In order to compare these results with the calibrated model, the failure stresses are normalized against the respective uniaxial strengths $\sigma_i^0$ with $i=$2,3. As can be noted, the model shows an excellent agreement with experimental data, a remarkable result considering that the spectral stiffness microplane model does not require any additional parameter to seize the multi-axial behavior of the material. As a reference, the Tsai-Wu failure envelope \cite{TsaWu72}, calibrated using only the uniaxial strengths in direction 2 and 3, is also plotted. As can be seen, the predictions from the microplane model and the Tsai-Wu criterion are very similar, with the microplane model being slightly more accurate especially in the II quadrant.

The behavior of the material under tension-torsional loading is represented in Figure (\ref{f10}b) which reports the experimental data by Fujii \emph{et. al} \cite{FujAmiLin92} on tubes made of a polyester matrix reinforced by a glass plain woven fabric. It can be seen that the model compares satisfactorily with the experimental data even if it tends to slightly overpredict the resistance to failure. The figure also shows the prediction provided by the Tsai-Wu criterion calibrated only with uniaxial data.

\subsubsection{Size effect tests}\label{size effect}
The fracture energy that was indirectly measured from size effect tests was incorporated in the model by calibrating the material point response. \emph{In-lieu} of the crack band model, the calibration satisfies the condition that the area under the stress-strain curve equals the fracture energy per unit volume (Figure \ref{f11}b). However, this does not guarantee that the model will be able to predict the right size effect in structural strength. This is because, for lab-scale structures of quasibrittle materials, the fracture process zone (FPZ) might not have reached its full size when the structure reached its peak load. Therefore, as was shown by Cusatis \emph{et al.} \cite{CusSch09}, the peak load is governed by not only the numerical value of the fracture energy, but also by the shape of the softening law. Thus, checking the predicted size effect is an important verification of the damage formulation. Material failure theories that are based purely on strength such as e.g. Tsai-Wu \cite{TsaWu72} cannot make this prediction. This aspect will be considered in detail in a separate
forthcoming article \cite{SalKirBaz15}. Here, only the comparison between the predicted size effect and the available test data is shown. Further, the Size Effect Law (SEL) proposed by \B~ \cite{BazPla98} fitting the experimental data is reported as a reference.

 Finite element models for all three SENT coupons were built and uniaxial tensile simulations were performed to predict the peak load. As noted in section \ref{crackband}, the fracture propagation for these specimens was modeled in the sense of the crack band model. The elements in the crack band had a height $h_e$ of 2 mm whereas the size in the direction of crack growth was 0.2 mm to capture correctly the stress profile in the ligament. Figure (\ref{f11}a) shows, for the coupons of different size, a comparison of the predicted and measured strengths, defined as $\sigma_N=P/tD$ with $P=$maximum load, $t=$thickness and $D=$witdh of the specimen. The calibrated material point stress-strain curve, instead, is depicted in Figure (\ref{f11}b). It can be seen that, indeed, the model predicts the correct size effect. It should be noted that the specific assumed form of the fiber softening is instrumental in this prediction. Furthermore, using the size effect fitting, the total fracture energy, $G_F$, estimated from the modeling predictions was 88.4 N/mm, which agrees very well with the value of the initial fracture energy, $G_f$, estimated from Size Effect Law ($\approx$ 74 N/mm \cite{SalKirBaz15}). In the absence of experimental data, a similar value was assumed in compression.

 The capability of capturing intra-laminar size effect of textile composites is key for safe damage tolerance design across various engineering applications. This is definitively a feature of the spectral stiffness microplane model as the excellent agreement with experiments showed.

\section{Application to crushing of composite tubes: results and discussion}
Thanks to their outstanding energy absorption capability, textile composites are often used as substitute of steel and aluminium to enhance crashworthiness of structural components. However, in order to fully exploit the potential of these materials, material models to be used as design tools are key. For this reason, the predictive capability of the spectral stiffness microplane model was challenged to predict the energy absorbed during the impact of composite crash cans. Figure (\ref{f11}c) shows the geometry of the structure under study consisting of a hat section tube and a reinforcing plate glued together by a toughened epoxy glue. Two different lay-up configurations were studied, namely:
\begin{enumerate}
  \ii  hat section tube: $[0^{\circ}]_{11}$, plate: $[0^{\circ}]_{8}$;
  \ii  hat section tube and plate: $[0^{\circ}/90^{\circ}/45^{\circ}/-45^{\circ}/0^{\circ}/90^{\circ}/0^{\circ}/-45^{\circ}/45^{\circ}/90^{\circ}/0^{\circ}]$;
\end{enumerate}

Following the test conditions described in section \ref{drop}, the crush cans were modeled in Abaqus Explicit \cite{Abaqus11} using a mesh of triangular shell elements of 2 mm. All the degrees of freedom of the nodes at the bottom section were fixed while a predefined velocity field of 4.6 m/s was prescribed to the impacting mass of 74.4 kg consisting of rigid shell elements. The general contact algorithm provided by Abaqus Explicit \cite{Abaqus11} was used while element deletion was adopted to avoid excessive element distortion during the simulation. The elements were deleted as soon as the dissipated energy reached $99\%$ of fracture energy or if the magnitude of the maximum or minimum principal strains reached 0.45.

The comparison between experimental and numerical results for case 1) and 2) is shown in Figures (\ref{f11}a,b) in terms of reaction force on the plate versus time while Figure (\ref{f11}d) shows the typical predicted fracturing pattern for case 1). As can be noted, a satisfactory agreement is found for both cases. As shown in Figure (\ref{f11}a) for case 1), the predicted plateau load, i.e. the reaction force on the plate once the crushing process is stabilized, compares very well with the experimental value: 33.7 kN vs 35.6 kN. However, not the same accuracy is provided for the time required to stop the mass, which is slightly overpredicted. This is not surprising since, as can be noted from Figure (\ref{f11}a) the reaction force at the early stages of the simulation is underpredicted. This might be an artifact of element deletion in the first stages of contact between the plate and the crush can and it is not considered here related to the material model. This was confirmed by the fact that running the same simulation with different material models provided the same problem in the early stages of the simulation. It is worth mentioning here that, during the tests, only 1/4 of the crush can was crashed, being the initial velocity not high enough. With an higher initial velocity, the prediction of the model in terms of time to stop the impacting mass would have been far more accurate. This because the extent of the region in which progressive stable crushing was present (predicted by the model with an error of about 5$\%$) would have been much larger.

Similarly, Figure (\ref{f11}b) shows the results for case 2). In this configuration, several layers of the material were loaded off-axis dissipating energy by extensive matrix microcracking before the onset of the first intra-laminar cracks. This damage mechanisms is supposed to provide extra energy dissipation compared to case 1) and, as a matter of facts, the experimental plateau load was in this case higher: 41.0 kN. Thanks to the excellent capability of capturing matrix microcracking, as was shown in the previous sections, the spectral stiffness microplane model was able to predict this increased force with a value of 42.6 kN. This proves that the model can be used as design tool to find the optimum lay-up configuration to maximize energy dissipation. It should also be noted that, in this case, also the total time to stop the impacting mass was predicted with very high accuracy. This may be in part due to the fact that the initial underprediction of the reaction force in the early stages of simulation was compensated by a slightly overpredicted plateau force.

It should be highlighted here that the results shown in Figures (\ref{f11}a-d) represent a pure prediction based on the calibration and validation done through uniaxial tests as well as size effect tests only. No adjustment of any of the parameters of the model was done, making the results shown even more remarkable. This achievement was made possible by the accurate modeling of the main damaging mechanisms such as e.g. matrix microcraking and, most importantly, by the correct modeling of intra-laminar cracking through the introduction in the material model of a characteristic length scale. The latter mechanism of energy dissipation seems to be the most important for the range of impact velocities considered in the present work, in which the maximum strain rate is not supposed to be higher than 20 s$^{-1}$. As reported from experiments, delamination seemed to be rather limited.

\section{Conclusions}
This contribution proposes a theoretical framework, called ``Spectral Stiffness Microplane Model", to simulate the orthotropic stiffness, pre-peak nonlinearity, failure envelopes, and the post-peak softening and fracture of textile composites.
Based on the results presented in this study, the following conclusions can be formulated:

\be  \setlength{\itemsep}{1.3mm}

\ii The microplane formulation with constitutive laws defined on planes of several orientations within the mesostructure allows a sound and physically-based description of the damage mechanisms occurring in textile composites whereas the spectral decomposition of the microplane strains and stresses provides a rigorous generalization of the formulation to anisotropy;

\ii Each eigenmode can be easily associated to the mechanical behavior of a particular constituent at the mesostructure and to a particular type of deformation. This makes it easy to take advantage of each eigenmode to define constitutive laws at microplane level targeting different damaging and fracturing mechanisms;

\ii Applied to a carbon twill 2$\times$2 composite, the model showed excellent agreement with uniaxial tests in tension and compression for various lay-ups. In particular, the model captured the highly non-linear mechanical behavior under off-axis and shear loading, which is typical of textile composites and is characterized by diffuse subcritical microcracking of the polymer matrix. This feature is of utmost importance in all situations in which predicting energy absorption is key, such as in crashworthiness analyses;

\ii Thanks to the use of microplanes representing the mesostructure of the material, the model captures the complex multi-axial behavior of textile composites without the need of any additional parameter;

\ii Different from strength-based criteria abundant in the literature, the formulation is endowed with a characteristic length through coupling with the crack band model. This ensures objective numerical analysis of softening damage and prevents spurious mesh sensitivity. Further, this is key to capture the intra-laminar size effect, a salient feature of composite structures. This aspect, too often overlooked in the literature on composites, is a determinant factor for damage tolerance design of large composite structures;

\ii  Compared with experimental results on the axial progressive crushing behavior of composite crush cans, the model provided excellent predictions for all the lay-up configurations under study. The reason is that the main damage mechanisms such as \emph{matrix microcracking} and longitudinal \emph{intra-laminar cracking} were captured correctly by the formulation;

\ii The model is computationally efficient and capable of analyzing the fracturing behavior of large composite structures, making it a valuable design tool for crashworthiness applications. Further, it has sufficient generality to allow extensions to composites with more complex architectures, such as the hybrid woven composites, and 2D or 3D woven or braided composites;

\ii The presented model is currently implemented in the commercial code MARS with the name ``Woven Composite Lamina".
 \ee
\section*{Acknowledgments}
This material is based upon work supported by the Department of Energy under Cooperative Award Number DE-EE0005661 to the United States Automotive Materials Partnership, LLC and sub-award SP0020579 to Northwestern University.
The work was also partially supported under NSF grant No. CMMI-1435923.




\clearpage
\listoftables
\listoffigures  
\clearpage
\begin{table}[ht]
\centering  
\begin{tabular}{l c c} 
 \hline
  \rule{0pt}{4ex}
 Description & Symbol (units) & Measured value\\[1 ex]
 \hline
In-plane modulus & $E$=$E_{2}$=$E_{3}$ (GPa)  & 53.5\\
Out-of-plane modulus & $E^{'}$ = $E_{1}$ (GPa) & 11.0$^{\mbox{a}}$ \\
In-plane shear modulus & $G$ = $G_{23}$ (GPa)  & 4.5 \\
Out-of-plane shear modulus & $G^{'}$ = $G_{12}$ = $G_{13}$ (GPa) & 3.6$^{\mbox{a}}$\\
In-plane Poisson ratio & $\nu=$$\nu_{23}$ = $\nu_{32}$ (-) & 0.055  \\
Out-of-plane Poisson ratio & $\nu^{'}$ = $\nu_{31}=$$\nu_{21}$ (-) & 0.4$^{\mbox{a}}$ \\
\hline
\multicolumn{3}{l}{$^{\mbox{a}}$properties estimated from \cite{SodHinKad98,SodHinKad02}.}
\end{tabular}
\caption{\sf Experimental elastic properties of carbon twill 2$\times$2 composite}
\label{T2}
\end{table}

\begin{table}
\scriptsize
\begin{tabular} { c c c  }
 \hline
  \rule{0pt}{4ex}
 Mode &$ \varepsilon_N^{(I)}$ &$ \varepsilon_T^{(I)}$\\[1 ex]
 \hline
 \rule{0pt}{4ex}
 $1$ &   $\alpha \left[\cos^2\theta+\left(\chi\cos^2\varphi+\sin^2\varphi\right)\sin^2\theta\right]$  & $\frac 1 2 \sqrt{\alpha^2\left(\chi-1\right)^2\cos^2\varphi\sin^2\theta\left(3-\cos2\varphi+2\cos^2\varphi\cos 2\theta\right)}$   \\
 $2$ &$\beta \left(\sin^2\theta-\sin^2\varphi\sin^2\theta\right)$ & $\sqrt{\beta^2\sin^2\theta\left[\sin^2 2\varphi + \cos^2\theta \left(\cos 2\varphi -3\right)^2\right]}$\\
 $3$    &$\gamma \left[\cos^2\theta+\left(\zeta\cos^2\varphi+\sin^2\varphi\right)\sin^2\theta\right]$ & $\frac 1 2 \sqrt{\gamma^2\left(\zeta-1\right)^2\cos^2\varphi\sin^2\theta\left(3-\cos2\varphi+2\cos^2\varphi\cos 2\theta\right)}$\\
 $4$&  $\varepsilon_4 \sin\varphi \sin2\theta$  & $\sqrt{\varepsilon_4^2 \left(\cos^2 \varphi\cos^2\theta+\cos^2 2\theta\sin^2\varphi\right)}$\\
 $5$& $\varepsilon_5 \cos\varphi\sin2\theta + \varepsilon_6 \sin^2\theta\sin 2\varphi$  & $\sqrt{\left(\varepsilon_5\cos\theta\sin\varphi-\varepsilon_6\cos 2\varphi\sin\theta\right)^2+\cos^2\varphi\left(\varepsilon_5\cos 2\theta+\varepsilon_6\sin\varphi\sin 2\theta\right)^2}$   \\ [1 ex]
 \hline
 \multicolumn{3}{l}{where $\alpha=\left(\varepsilon_2+\varepsilon_3+\varepsilon_1 \chi\right)/\left(2+\chi^2\right)$, $\beta=1/2\left(\varepsilon_3-\varepsilon_2\right)$,
$\gamma=\left(\varepsilon_2+\varepsilon_3+\varepsilon_1 \zeta\right)/\left(2+\zeta^2\right)$}
\end{tabular}
\caption{\sf Microplane eigenstrains: normal and tangential components}
\label{T1}
\end{table}

\begin{table}[ht]
\scriptsize

\centering  
\begin{tabular}{c l c c} 
 \hline
  \rule{0pt}{4ex}
 Mode &Description & Symbol (units) & Calibrated value\\[1 ex]
 \hline
\multirow{8}{*} {12} & mode 1 elastic eigenvalue & $\lambda^{(1)}$ (GPa) & 61.85$^{\mbox{a}}$\\
& mode 2 elastic eigenvalue & $\lambda^{(2)}$ (GPa) & 50.71$^{\mbox{a}}$\\
& microplane peak stress in tension & $s_0^{(12)}$ (MPa) & 400\\
& parameter governing post-peak softening in tension  & $k_{bt}^{(12)}$ (-) & $30.60\times10^{-3}$\\
& parameter governing post-peak softening in tension  & $a_{12t}$ (-) & 0.75\\
& microplane peak stress in compression & $c_0^{(12)}$ (MPa) & 405\\
& parameter governing post-peak softening in compression  & $k_{bc}^{(12)}$ (-) & $30.60\times10^{-3}$\\
& parameter governing post-peak softening in compression  & $a_{c12}$ (-) & 0.75\\[1 ex]
\hline
\multirow{8}{*} {4} & mode 4 elastic eigenvalue & $\lambda^{(4)}$ (GPa) & 8.10$^{\mbox{a}}$\\
& microplane stress in tension at start of non-linear boundary & $s_0^{(4)}$ (MPa) & 45\\
& exponent governing pre-peak non-linearity in tension and compression  & $p$ (-) & $0.3$\\
& strain at starting of post-peak softening in tension  & $k_{at}^{(4)}$ (-) & $124.6\times10^{-3}$\\
& parameter governing post-peak softening in tension  & $k_{bt}^{(4)}$ (-) & $120.15\times10^{-3}$\\
& microplane stress in compression at start of non-linear boundary & $c_0^{(4)}$ (MPa) & 45\\
& strain at starting of post-peak softening in compression  & $k_{ac}^{(4)}$ (-) & $124.6\times10^{-3}$\\
& parameter governing post-peak softening in compression  & $k_{bc}^{(4)}$ (-) & $120.15\times10^{-3}$\\[1 ex]

\hline
\multicolumn{4}{l}{$^{\mbox{a}}$ Calculated by means of Eqs. (\ref{e8}a-d).}
\end{tabular}
\caption{\sf Material model parameters, calibrated by means of uniaxial and size effect tests, describing the in-plane behavior of the composite.}
\label{T3}
\end{table}

\begin{table}[ht]
\scriptsize

\centering  
\begin{tabular}{c l c c} 
 \hline
  \rule{0pt}{4ex}
 Mode &Description & Symbol (units) & Calibrated value\\[1 ex]
 \hline
\multirow{7}{*} {3} & mode 3 elastic eigenvalue & $\lambda^{(3)}$ (GPa) & 10.82$^{\mbox{a}}$\\
& microplane peak stress in tension & $s_0^{(3)}$ (MPa) & 90\\
& strain at starting of post-peak softening in tension  & $k_{at}^{(3)}$ (-) & $4.0\times10^{-3}$\\
& parameter governing post-peak softening in tension  & $k_{bt}^{(3)}$ (-) & $20\times10^{-3}$\\
& microplane peak stress in compression & $c_0^{(3)}$ (MPa) & 90\\
& strain at starting of post-peak softening in compression  & $k_{ac}^{(3)}$ (-) & $4.0\times10^{-3}$\\
& parameter governing post-peak softening in compression  & $k_{bc}^{(3)}$ (-) & $20\times10^{-3}$\\[1 ex]
\hline
\multirow{8}{*} {5} & mode 5 elastic eigenvalue & $\lambda^{(5)}$ (GPa) & 7.20$^{\mbox{a}}$\\
& microplane stress in tension at start of non-linear boundary & $s_0^{(5)}$ (MPa) & 45\\
& exponent governing pre-peak non-linearity in tension and compression  & $p$ (-) & $0.3$\\
& strain at starting of post-peak softening in tension  & $k_{at}^{(5)}$ (-) & $124.6\times10^{-3}$\\
& parameter governing post-peak softening in tension  & $k_{bt}^{(5)}$ (-) & $120.15\times10^{-3}$\\
& microplane stress in compression at start of non-linear boundary & $c_0^{(5)}$ (MPa) & 45\\
& strain at starting of post-peak softening in compression  & $k_{ac}^{(5)}$ (-) & $124.6\times10^{-3}$\\
& parameter governing post-peak softening in compression  & $k_{bc}^{(5)}$ (-) & $120.15\times10^{-3}$\\[1 ex]
\hline
\multicolumn{4}{l}{$^{\mbox{a}}$ Calculated by means of Eqs. (\ref{e8}a-d).}
\end{tabular}
\caption{\sf Material model parameters related to the out-of-plane behavior of the composite. The parameters for mode 3 were estimated from \cite{SodHinKad98,SodHinKad02}. Mode 5 parameters were assumed to have the same values as for mode 4.}
\label{T4}
\end{table}

\clearpage
\bfi \center
  \includegraphics[trim=0cm 0.5cm 0cm 1cm, clip=true,clip=true,width = 1\textwidth]{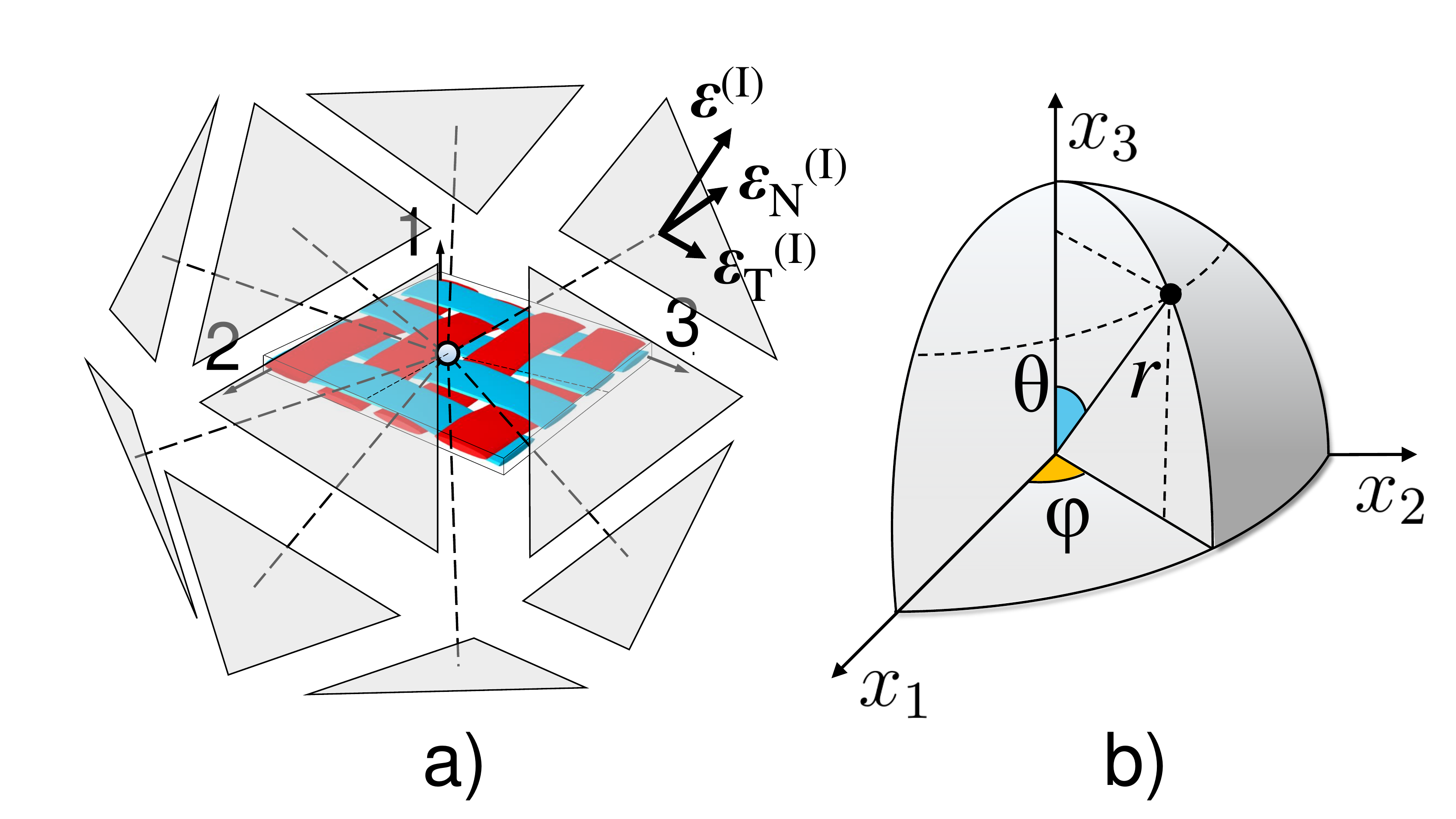} \caption{\label{f1} \sf Schematic representation of a) the Representative Unit Cell of a 2$\times$2 twill composite with its local coordinate system and the microplanes used to define the constitutive laws of the material; b) local spherical coordinate system.} \efi

\bfi \center
  \includegraphics[trim=0cm 0.9cm 0.5cm 1cm, clip=true,clip=true,width = 1.0\textwidth]{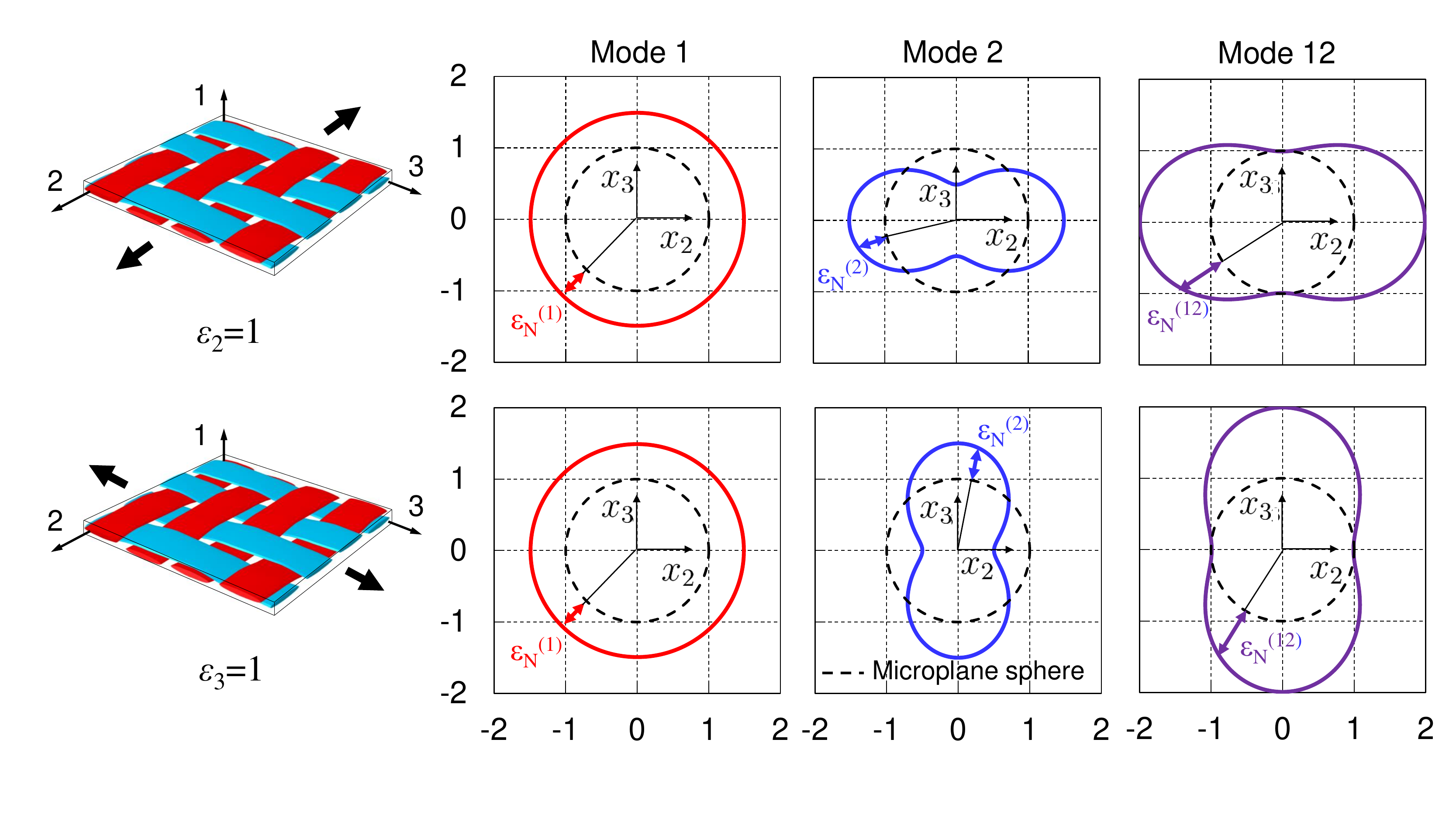} \caption{\label{f2} \sf Distribution of the normal microplane strain related to a) mode 1, b) mode 2 and c) mode12 in the presence of a uniaxial macroscopic strain in plane 2-3.} \efi
\bfi \center
  \includegraphics[trim=7cm 5.5cm 7cm 3cm, clip=true,clip=true,width = 1.0\textwidth]{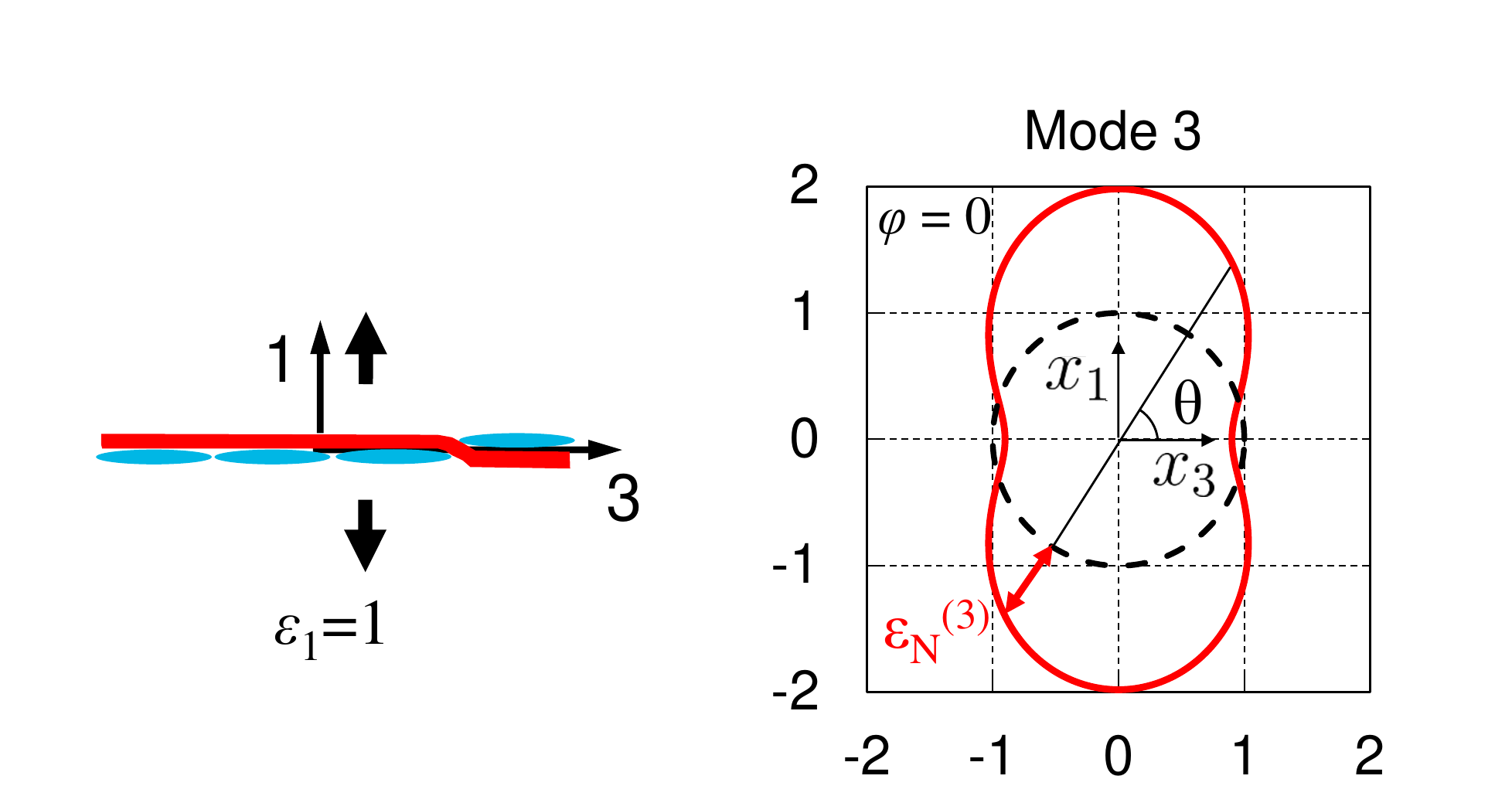} \caption{\label{f3} \sf Distribution of the normal microplane strain related to mode 3 in the presence of a uniaxial macroscopic strain $\varepsilon_1$.} \efi

\bfi \center
  \includegraphics[trim=7cm 5.5cm 7cm 3cm, clip=true,clip=true,width = 1.0\textwidth]{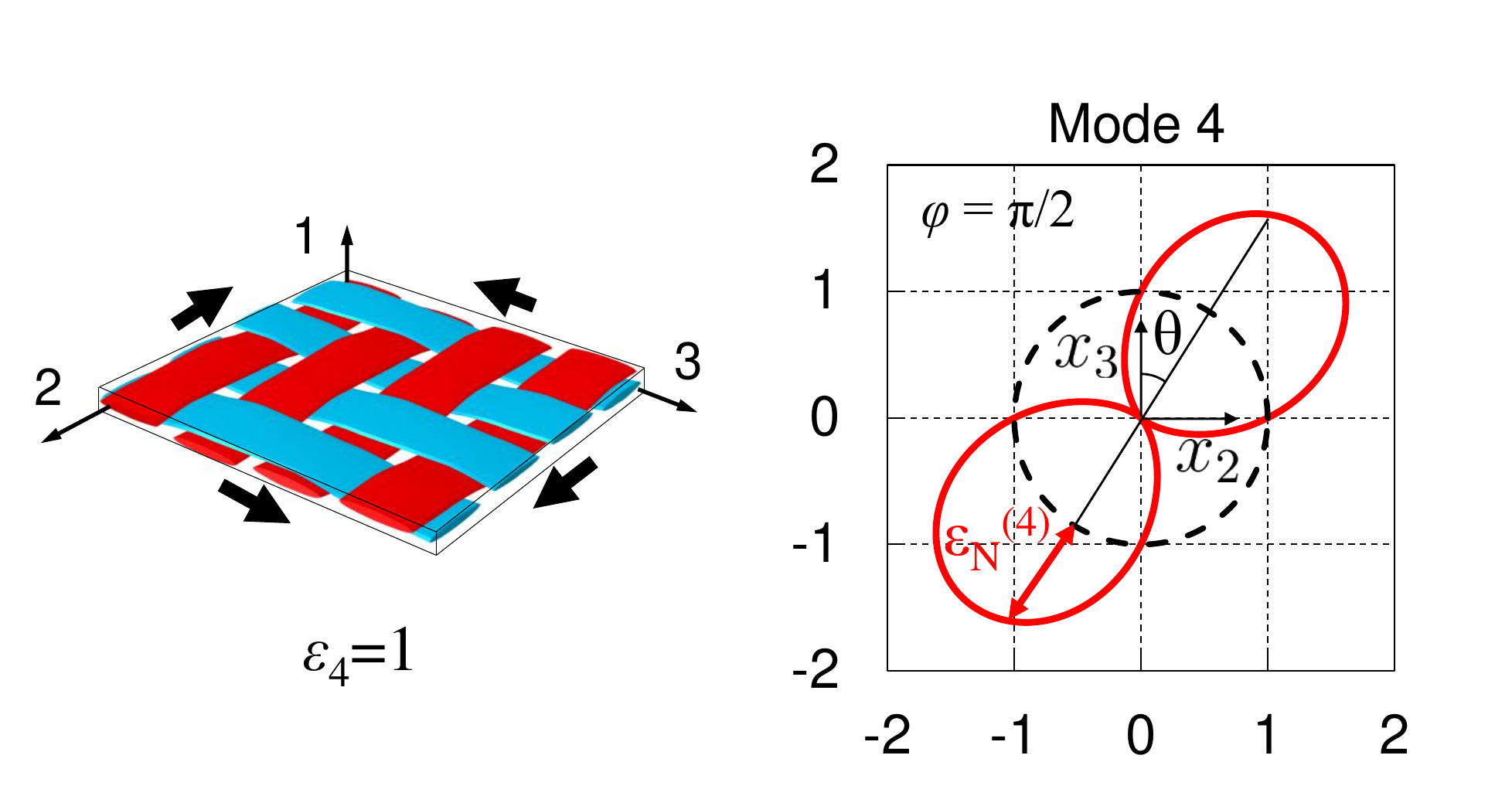} \caption{\label{f4} \sf Distribution of the normal microplane strain related to mode 4 in the presence of a macroscopic in-plane shear strain $\varepsilon_4$.} \efi

\bfi \center
  \includegraphics[trim=7cm 1.5cm 7cm 0.5cm, clip=true,clip=true,width = 1.0\textwidth]{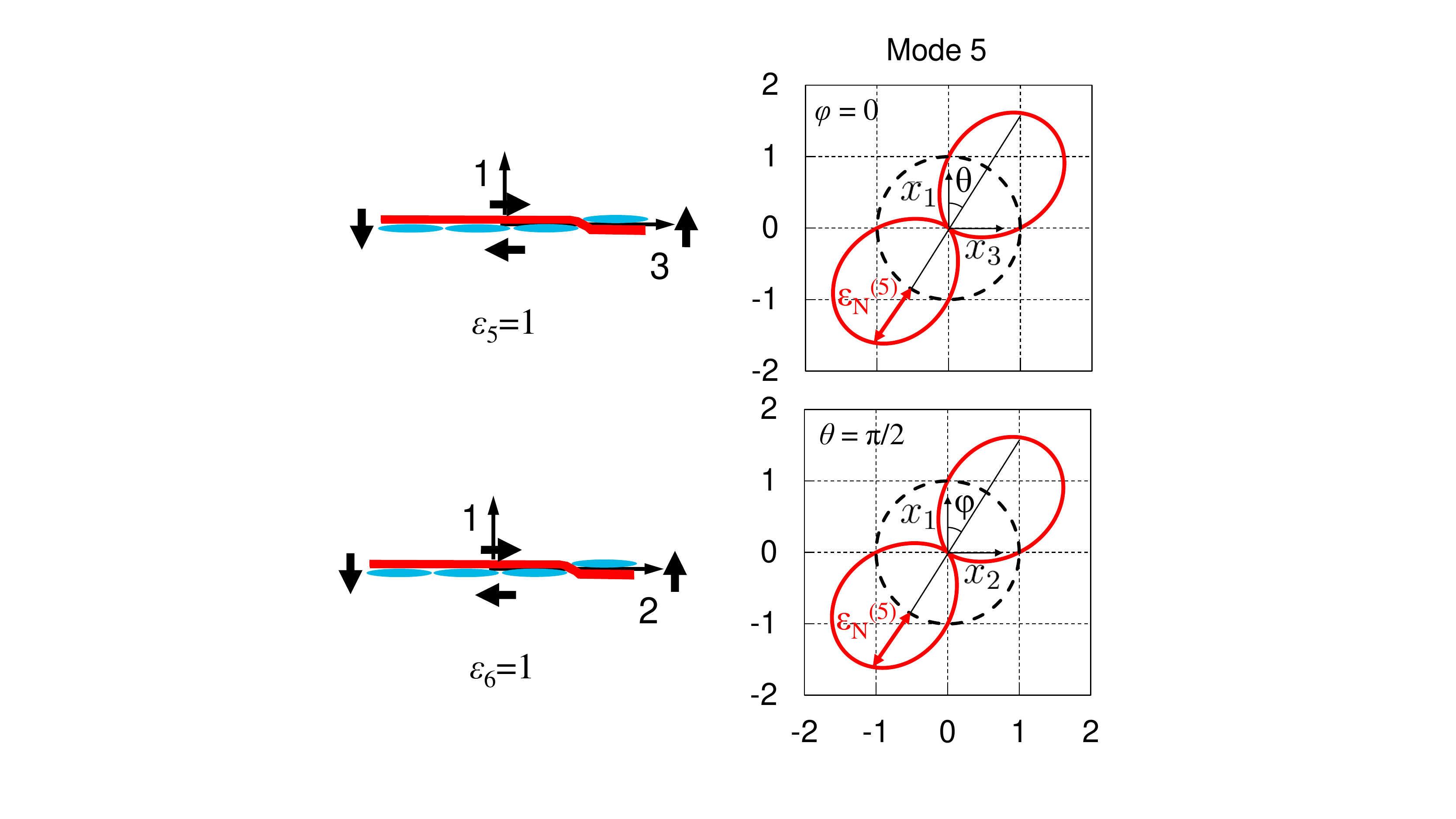} \caption{\label{f5} \sf Distribution of the normal microplane strain related to mode 5 in the presence of macroscopic out-of-plane shear strains $\varepsilon_5$ and $\varepsilon_6$.} \efi
\bfi \center
  \includegraphics[trim=3cm 0cm 3cm 0cm, clip=true,clip=true,width = 1.0\textwidth]{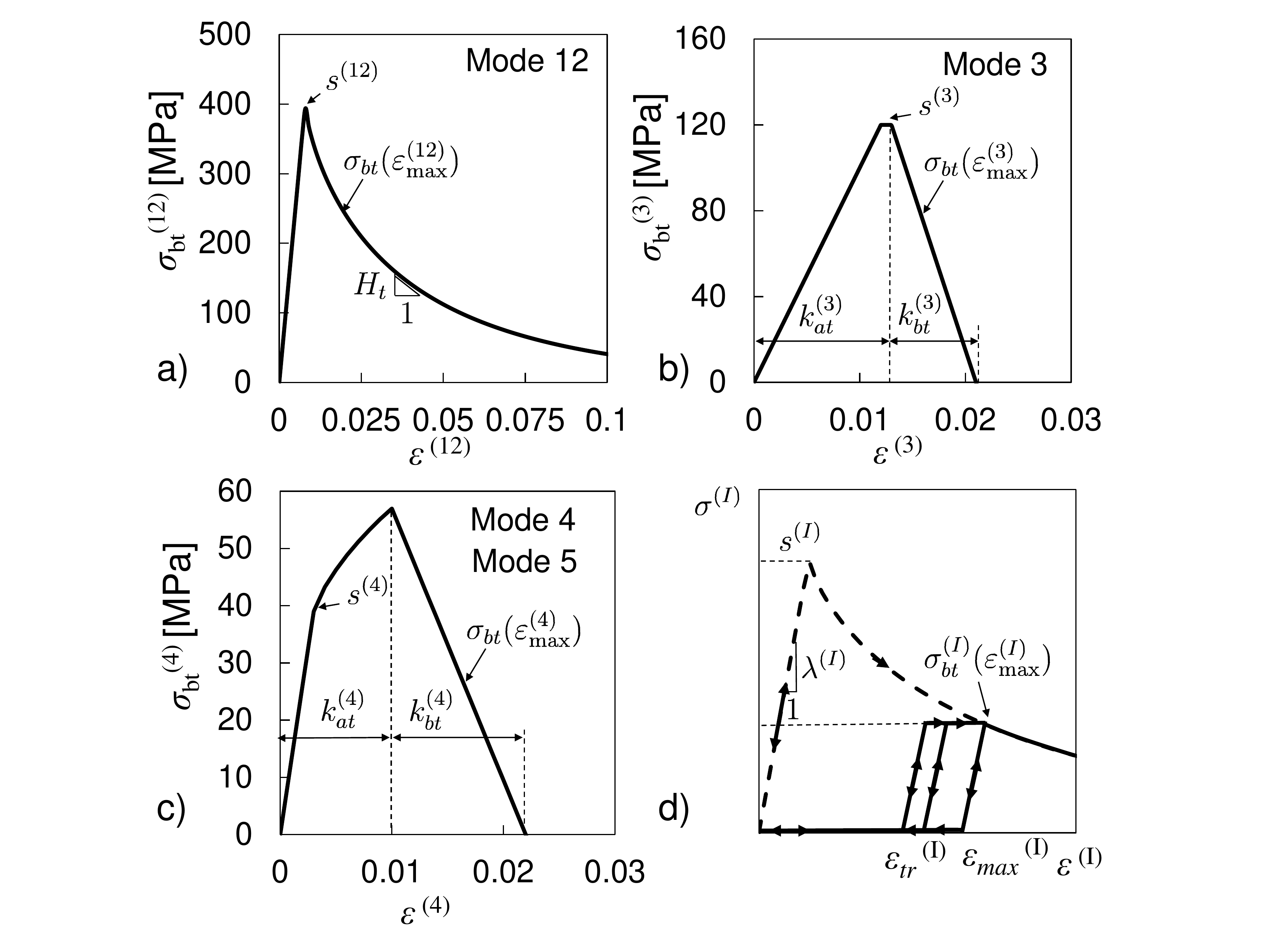} \caption{\label{f6} \sf Spectral stiffness microplane constitutive laws. a) Typical stress boundary for mode 12 with exponential decay; b) Typical bilinear boundary for mode 3; c) mode 4 and 5 inelastic stress boundaries; d) typical unloading-reloading rule for a $I$-th generic mode.} \efi
\bfi \center
  \includegraphics[trim=1cm 3.5cm 0.8cm 3.5cm, clip=true,clip=true,width = 1.0\textwidth]{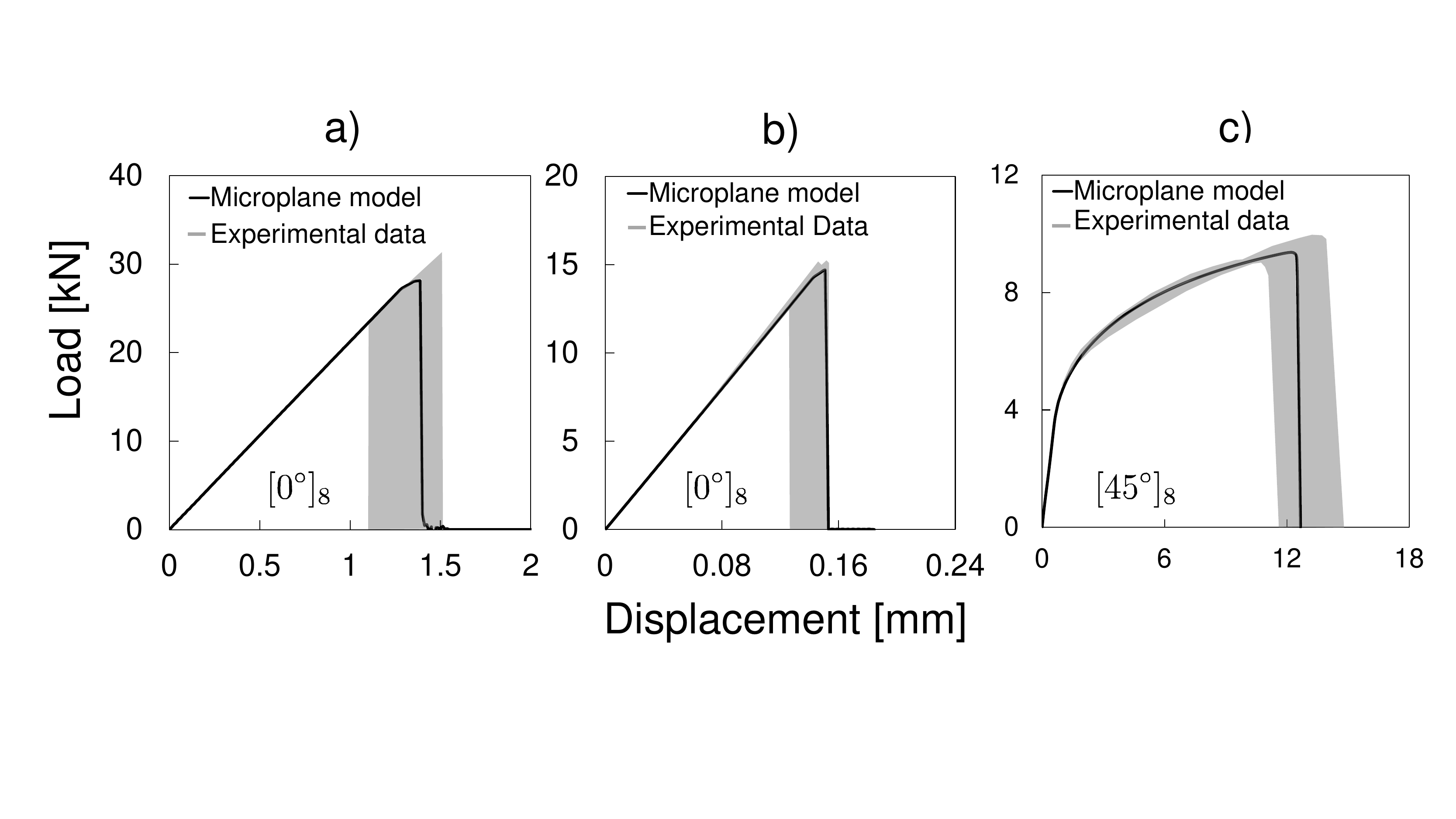} \caption{\label{f7} \sf Comparison of the measured and calibrated load displacement curve for a) a $[0]_8$ coupon, under tension; b) a $[0]_8$ coupon, under compression and c) a $[45]_8$ coupon, under tension. } \efi
\bfi \center
  \includegraphics[trim=1.5cm 2.8cm 2cm 1cm, clip=true,clip=true,width = 1.0\textwidth]{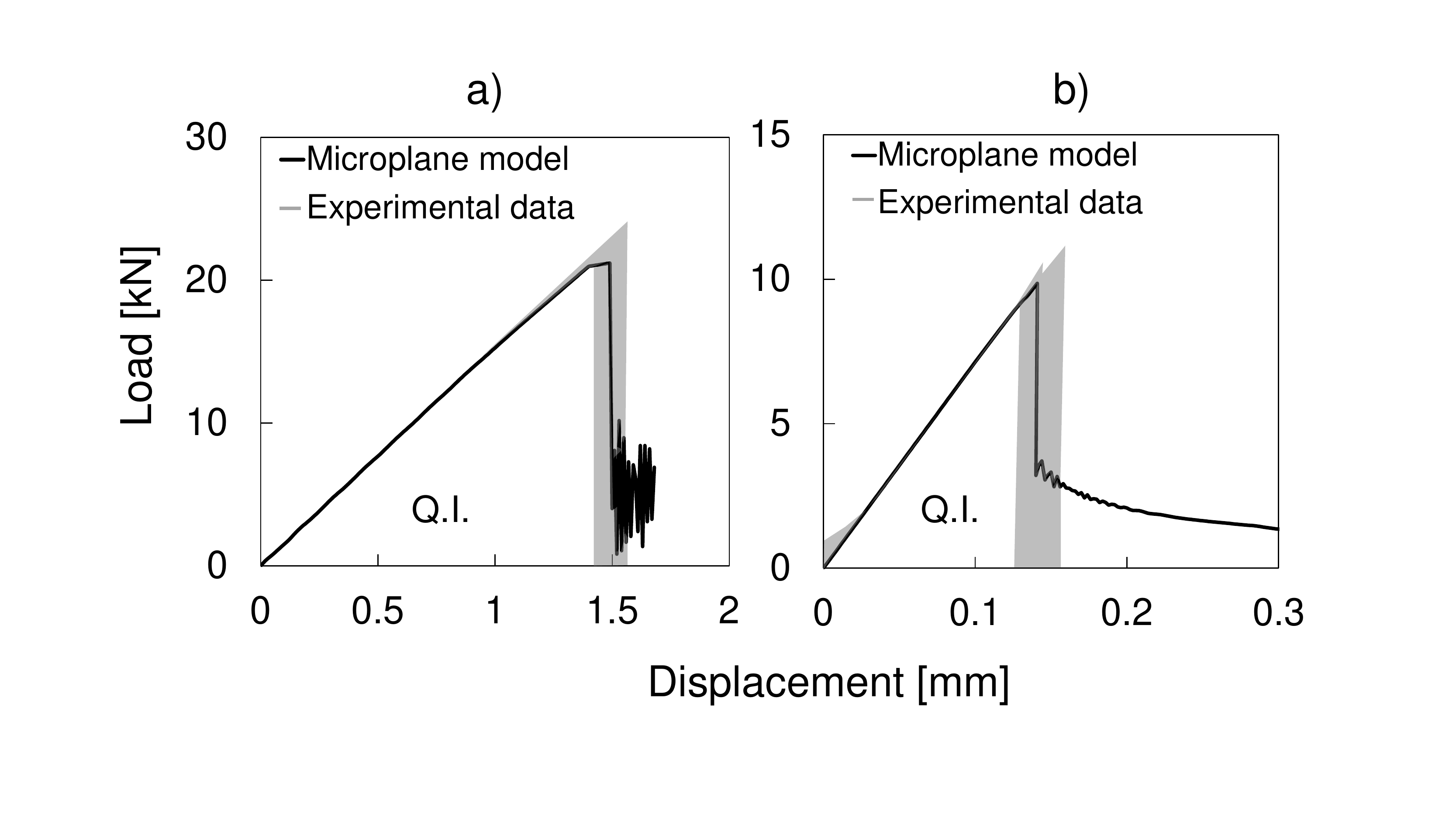} \caption{\label{f8} \sf Comparison of the measured and predicted load displacement curve for a quasi-isotropic coupon $[0^{\circ}/45^{\circ}/-45^{\circ}/90^{\circ}$]$_{2s}$ under (a) tension, (b) compression.} \efi
\bfi \center
  \includegraphics[trim=2.5cm 6.5cm 2cm 1cm, clip=true,clip=true,width = 1.0\textwidth]{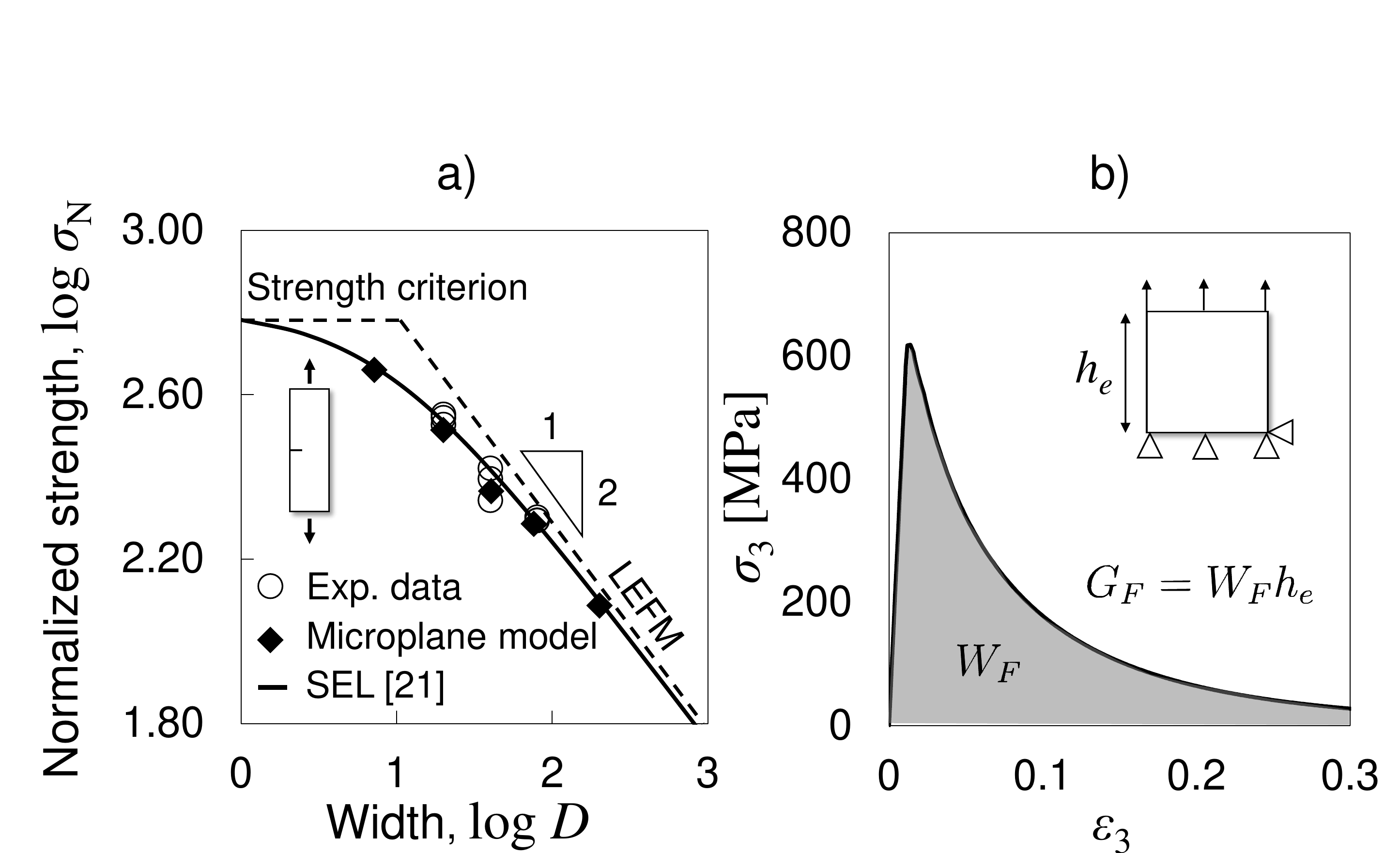} \caption{\label{f9} \sf a) Comparison of the measured and predicted size effect in geometrically scaled Single Edge Notch Tension $[0]_8$ coupons; b) typical stress-strain curve in pure tension. The total fracture energy $G_F$ is calibrated adjusting the post-peak softening response of the material.} \efi

\bfi \center
  \includegraphics[trim=0cm 4.8cm 0cm 1.5cm, clip=true,clip=true,width = 1.0\textwidth]{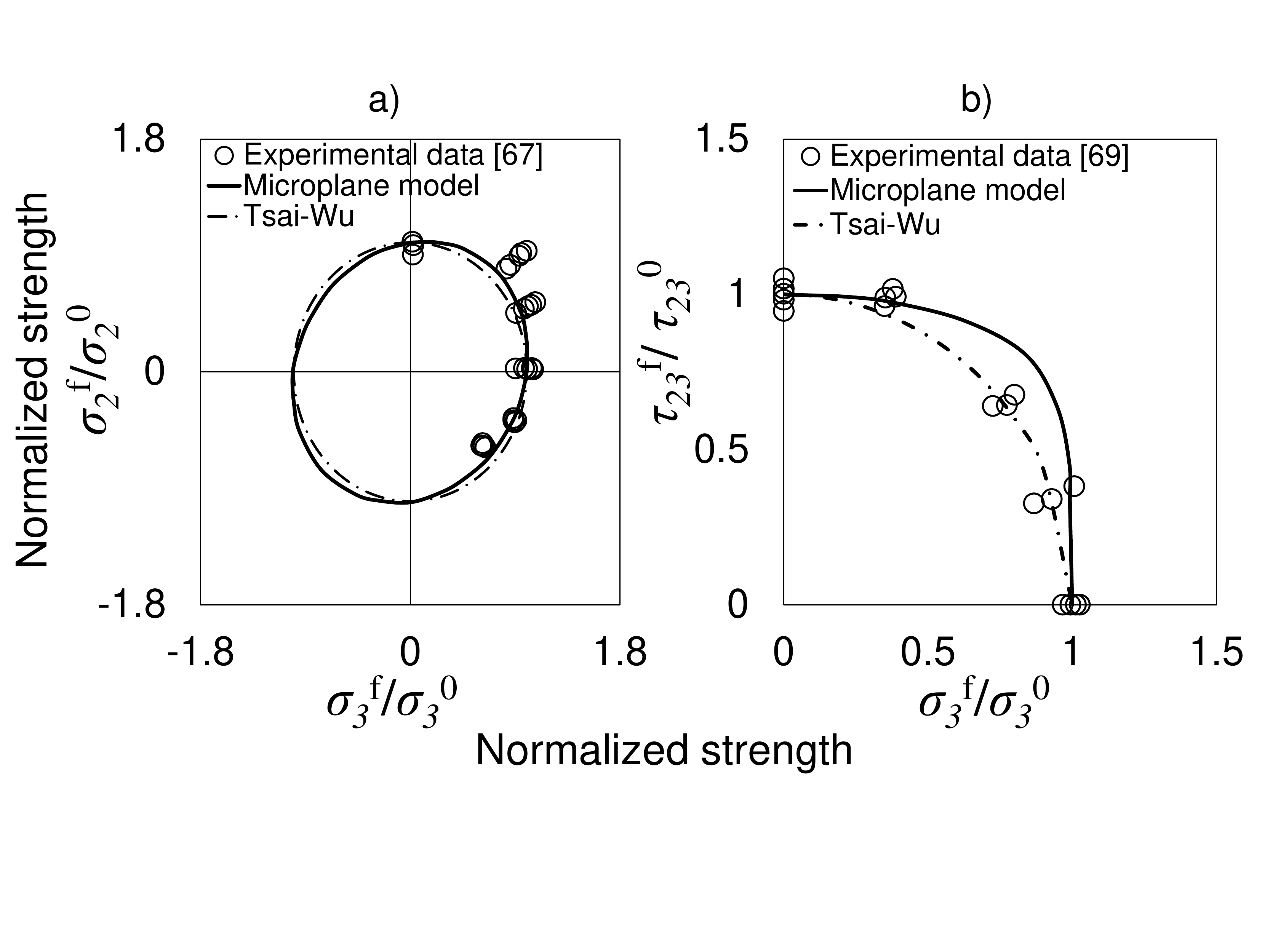} \caption{\label{f10} \sf Comparison between measured and predicted biaxial behavior. a) Biaxial loading; b) tension-torsion loading.} \efi

\bfi \center
  \includegraphics[trim=0cm 3cm 0cm 1.5cm, clip=true,clip=true,width = 1.0\textwidth]{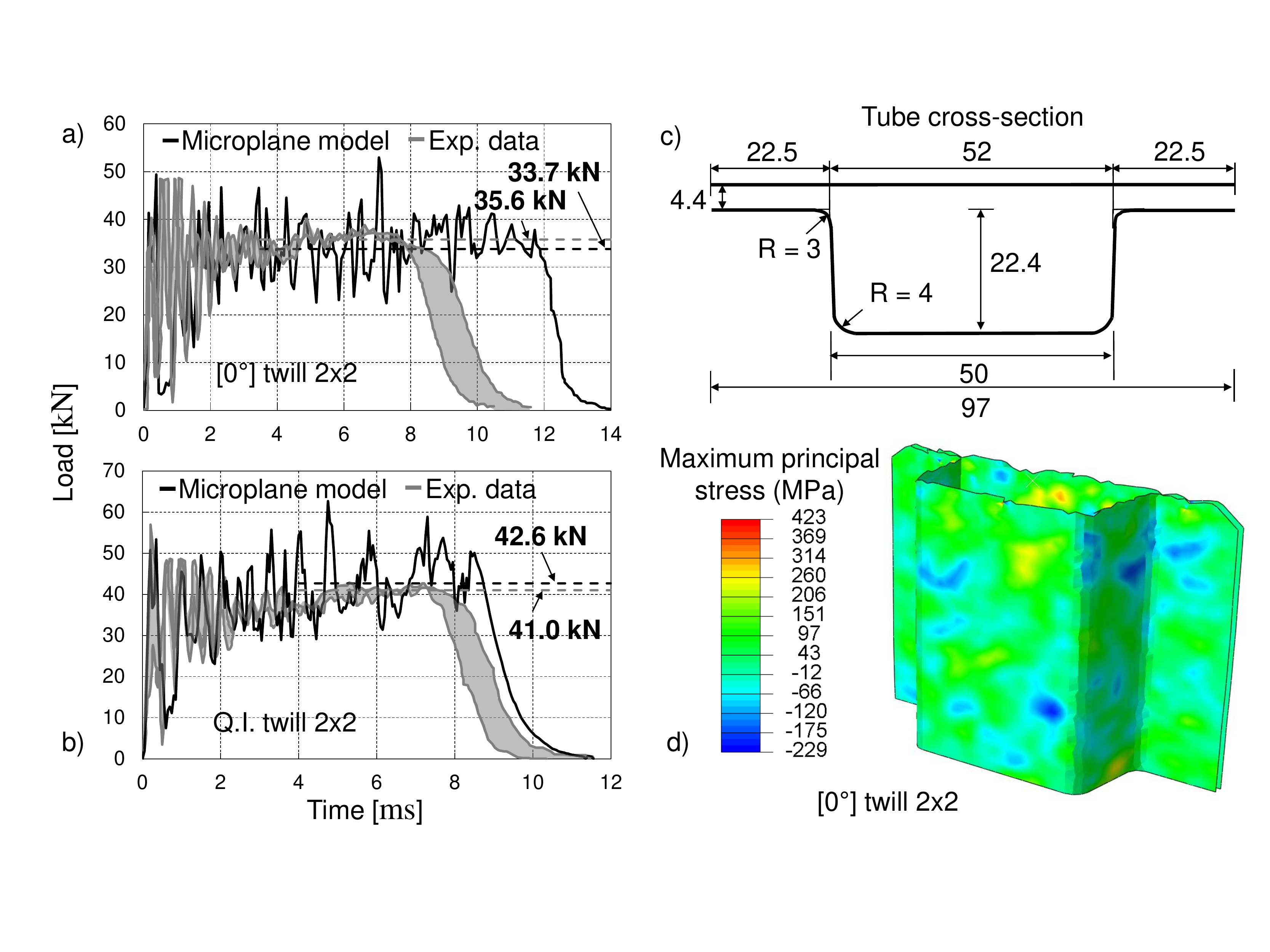} \caption{\label{f11} \sf Crashing of composite tubes. a) comparison between model prediction and experimental results (hat section: $[0^{\circ}]_{11}$, plate: $[0^{\circ}]_8$); b) comparison between model prediction and experimental results (hat section and plate: $[0^{\circ}/90^{\circ}/45^{\circ}/-45^{\circ}/0^{\circ}/90^{\circ}/0^{\circ}/-45^{\circ}/45^{\circ}/90^{\circ}/0^{\circ}]$); c) geometric specification of the crash can cross-section (dimensions in mm); d) typical predicted fracturing pattern.} \efi

\end{document}